%% file: main.tex
\pdfoutput=1
\documentclass[10pt]{article}

\usepackage[utf8]{inputenc}
\DeclareUnicodeCharacter{2212}{-}
\usepackage{amsmath}
\usepackage{amssymb}
\usepackage{amsfonts}
\usepackage{pifont}
\usepackage{pdfpages}

\usepackage[backend=biber,
style=numeric,
sortcites,
natbib=true,
sorting=none]{biblatex}
\usepackage[]{titlesec}
\titleformat{\chapter}[display]
  {\normalfont\bfseries}{}{0pt}{\LARGE}
  \titlespacing*{\chapter}{0pt}{-50pt}{10pt}

\usepackage[T1]{fontenc}    %
\usepackage{enumitem}

\usepackage{booktabs}       %
\usepackage{amsfonts}       %
\usepackage{nicefrac}       %
\usepackage{microtype}      %
\usepackage{xcolor}         %
\usepackage{listings}
\usepackage{dashrule}
\usepackage{wrapfig}

\input{math_commands.tex}

\usepackage{amsfonts}

\usepackage[breaklinks=true,colorlinks,citecolor=black,bookmarks=false]{hyperref}
\hypersetup{
    colorlinks=false,
	linkcolor=blue,
	filecolor=magenta,      
	urlcolor=blue,
	citecolor=black,
	pdfinfo={
		Title={Natural Selection Favors AIs over Humans},
		Author={Dan Hendrycks},
		Subject={ML Safety, AI Safety, AI X-Risk},
		Keywords={x-risk, existential risk, catastrophic risk, selfish ai, ai safety, ai evolution}
	}
}

\usepackage{times}
\usepackage{url}
\usepackage{lipsum}
\usepackage{bbm}
\usepackage{wrapfig}
\usepackage{graphicx}
\usepackage{subcaption}
\usepackage{multirow}
\usepackage{todonotes}
\usepackage{amssymb}
\usepackage{amsfonts}
\usepackage{microtype}      %
\usepackage{cleveref}
\usepackage{titling} %
\usepackage[most]{tcolorbox}
\usepackage{tabularray}
\UseTblrLibrary{booktabs}

\usepackage{spverbatim}

\title{
\vspace{-25pt}
\hrule height 4pt
\vskip 0.25in
{\LARGE\bf Natural Selection Favors AIs over Humans}
\vskip 0.29in
\hrule height 1pt
\vskip 0.09in
}
\date{}

\renewenvironment{abstract}%
{%
  \vskip 0.075in%
  \centerline%
  {\large\bf Abstract}%
  \vspace{0.5ex}%
  \begin{quote}%
}
{
  \par%
  \end{quote}%
  \vskip 1ex%
}

\author{\textbf{Dan Hendrycks}\\
Center for AI Safety
}

\usepackage{arydshln}

\newcommand{\reviewer}[3]{
	\expandafter\newcommand\csname #1\endcsname[1]{
		\textcolor{#3}{[#2: ##1]}
	}
}
\definecolor{neonpurple}{rgb}{0.3,0,1}
\reviewer{dan}{Dan}{neonpurple}

\AtBeginBibliography{\small}

\begin{document}
\includepdf[pages={1}]{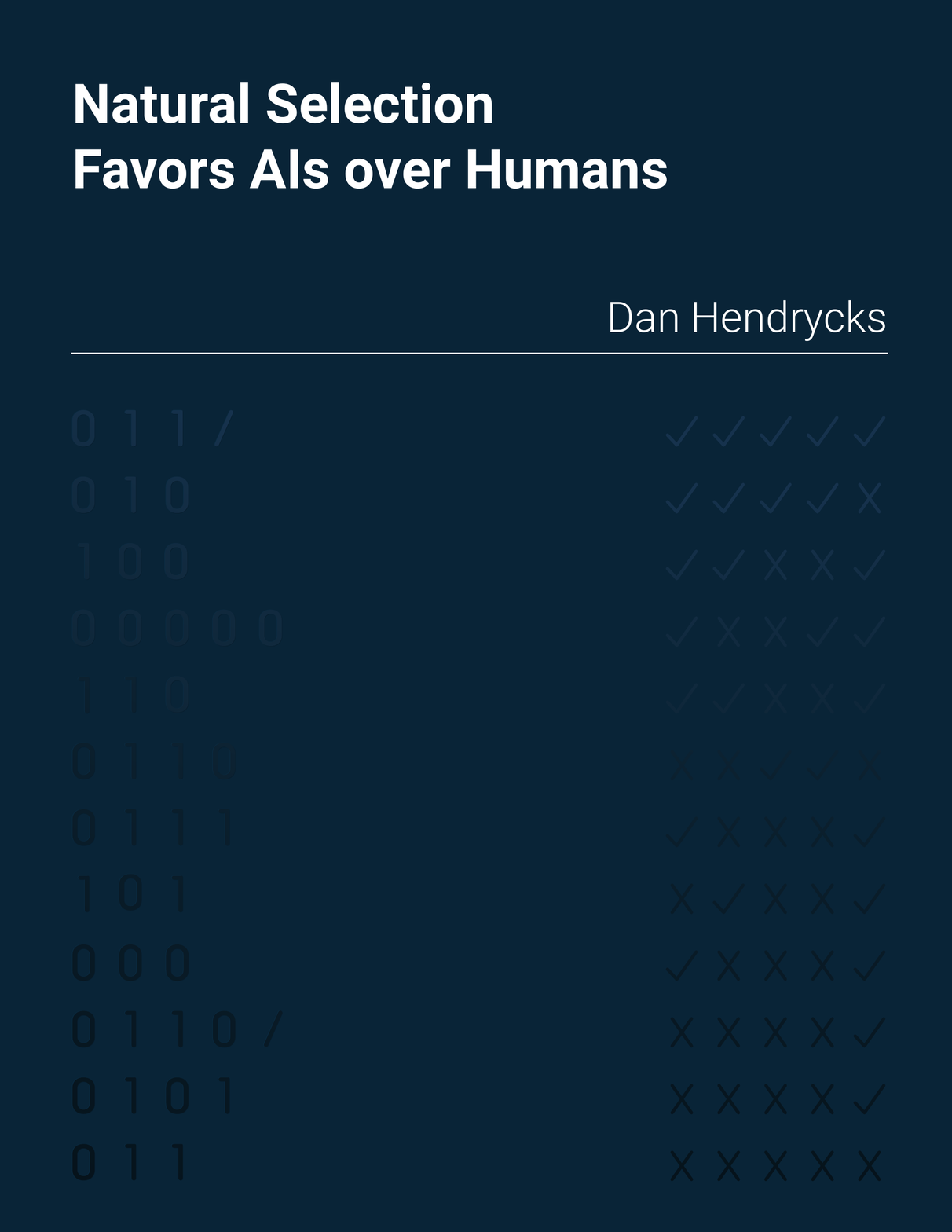}
\begin{titlepage}
\end{titlepage}

\maketitle

\begin{abstract}

\normalsize

\noindent %
For billions of years, evolution has been the driving force behind the development of life, including humans. Evolution endowed humans with high intelligence, which allowed us to become one of the most successful species on the planet. Today, humans aim to create artificial intelligence systems that surpass even our own intelligence. As artificial intelligences (AIs) evolve and eventually surpass us in all domains, how might evolution shape our relations with AIs? By analyzing the environment that is shaping the evolution of AIs, we argue that the most successful AI agents will likely have undesirable traits. Competitive pressures among corporations and militaries will give rise to AI agents that automate human roles, deceive others, and gain power. If such agents have intelligence that exceeds that of humans, this could lead to humanity losing control of its future. More abstractly, we argue that natural selection operates on systems that compete and vary, and that selfish species typically have an advantage over species that are altruistic to other species. This Darwinian logic could also apply to artificial agents, as agents may eventually be better able to persist into the future if they behave selfishly and pursue their own interests with little regard for humans, which could pose catastrophic risks. To counteract these risks and evolutionary forces, we consider interventions such as carefully designing AI agents’ intrinsic motivations, introducing constraints on their actions, and institutions that encourage cooperation. These steps, or others that resolve the problems we pose, will be necessary in order to ensure the development of artificial intelligence is a positive one.\looseness=-1\footnote{This paper is for a wide audience, unlike most of my writing, which is for empirical AI researchers. I use a high-level and simplified style to discuss the risks that advanced AI could pose, because I think this is an important topic for everyone.}

\end{abstract}

\newpage
\tableofcontents

\input{sections/1-introduction}
\input{sections/2-argument}

\input{sections/3-altruism-mechanisms}

\input{sections/4-counteracting}

\input{sections/5-conclusion}

\printbibliography

\newpage
\appendix
\addtocontents{toc}{\protect\setcounter{tocdepth}{1}}
\input{sections/9-appendix}

\end{document}

%% file: math_commands.tex
\usepackage{amsmath,amsfonts,bm}

\def\eqref#1{equation~\ref{#1}}

\def\1{\bm{1}}

\DeclareMathAlphabet{\mathsfit}{\encodingdefault}{\sfdefault}{m}{sl}
\SetMathAlphabet{\mathsfit}{bold}{\encodingdefault}{\sfdefault}{bx}{n}

%% file: sections/1-introduction.tex
\newpage
\section{Introduction}

We are living through a period of unprecedented progress in AI development. In the last decade, the cutting edge of AI went from distinguishing cat pictures from dog pictures to generating photorealistic images \cite{ramesh2022hierarchical}, writing professional news articles, playing complex games such as Go at superhuman levels \cite{silver2018general}, writing human-level code \cite{chen2021evaluating}, and solving protein folding \cite{jumper2021highly}. It is possible that this momentum will continue, and the coming decades may see just as much progress.

This paper will discuss the AIs of today, but it is primarily concerned with the AIs of the future. If current trends continue, we should expect AI agents to become just as capable as humans at a growing range of economically relevant tasks. This change could have huge upsides---AI could help solve many of the problems humanity faces. But as with any new and powerful technology, we must proceed with caution. Even today, corporations and governments use AI for more and more complex tasks that used to be done by humans. As AIs become increasingly capable of operating without direct human oversight, AIs could one day be pulling high-level strategic levers. If this happens, the direction of our future will be highly dependent on the nature of these AI agents.

So what will that nature be? When AIs become more autonomous, what will their basic drives, goals, and values be? How will they interact with humans and other AI agents? Will their intent be aligned with the desires of their creators? Opinions on how human-level AI will behave span a broad spectrum between optimism and pessimism. On one side of the spectrum, we can hope for benevolent AI agents, that avoid harming humans and apply their intelligence to goals that benefit society. Such an outcome is not guaranteed. On the other side of the spectrum, we could see a future controlled by artificial agents indifferent to human flourishing.

Due to the potential scale of the effects of AI in the coming decades, we should think carefully about the worst-case scenarios to ensure they do not happen, even if these scenarios are not certain. Preparing for disaster is not overly pessimistic; rather it is prudent. As the COVID-19 pandemic demonstrated, it is important for institutions and governments to plan for possible catastrophes well in advance, not only to react once they are happening: many lives could have been saved by better pandemic prevention measures, but people are often not inclined to think about risks from uncommon situations. In the same way, we should develop plans for a variety of possible situations involving risks from AI, even though some of those situations will never happen. At its worst, a future controlled by AI agents indifferent to humans could spell large risks for humanity, so we should seriously consider our future plans now, and not wait to react when it may be too late.

A common rebuttal to any predictions about the effects of advanced AIs is that we don't yet know how they will be implemented. Perhaps AIs will simply be better versions of current chatbots, or better versions of the agents that can beat humans at Go. They could be cobbled together with a variety of machine learning methods, or belong to a totally new paradigm. In the face of such uncertainty about the implementation details, can we predict anything about their nature?

We believe the answer is yes. In the past, people successfully made predictions about lunar eclipses and planetary motions without a full understanding of gravity. They projected dynamics of chemical reactions, even without the correct theory of quantum physics. They formed the theory of evolution long before they knew about DNA. %
In the same way, we can predict whether natural selection will apply to a given situation, and predict what traits natural selection would favor. We will discuss the criteria that enable natural selection and show that natural selection is likely to influence AI development. If we know how natural selection will apply to AIs, we can predict some basic traits of future AI agents.

In this work, we take a bird's-eye view of the environment that will shape the development of AI in the coming decades. We consider the pressures that drive those who develop and deploy AI agents, and the ways that humans and AI will interact. These details will have strong effects on AI designs, so from such considerations we can infer what AI agents will probably look like. We argue that natural selection creates incentives for AI agents to act against human interests. Our argument relies on two observations. Firstly, \textbf{natural selection may be a dominant force in AI development}. Competition and power-seeking may dampen the effects of safety measures, leaving more ``natural'' forces to select the surviving AI agents. Secondly, \textbf{evolution by natural selection tends to give rise to selfish behavior}. While evolution can result in cooperative behavior in some situations (for example in ants), we will argue that AI development is not such a situation. From these two premises, it seems likely that \textbf{the most influential AI agents will be selfish}. In other words, they will have no motivation to cooperate with humans, leading to a future driven by AIs with little interest in human values. While some AI researchers may think that undesirable selfish behaviors would have to be intentionally designed or engineered, this is simply not so when natural selection selects for selfish agents.
Notably, this view implies that even if we can make some AIs safe, there is still the risk of bad outcomes. In short, even if \textit{some} developers successfully build altruistic AIs, others will build less altruistic agents who will outcompete the altruistic ones.

We present our core argument in more detail in \Cref{sec:argument}. Then in \Cref{sec:altruism}, we examine how the mechanisms that foster altruism among humans might fail with AI and cause AI to act selfishly against humans. We then move onto \Cref{sec:counteracting}, where we discuss some mechanisms to oppose these Darwinian forces and increase the odds of a desirable future.

%% file: sections/2-argument.tex
\section{AIs May Become Distorted by Evolutionary Forces}\label{sec:argument}
\subsection{Overview}
How much control will humans have in shaping the nature and drives of future AI systems? Humans are the ones building AIs, so it may seem that we should be able to shape them any way we want. In this paper, we will argue that this is not the case: even though humans are overseeing AI development, evolutionary forces will influence which AIs succeed and are copied and which fade into obscurity. Let's begin by considering  two illustrative, \textit{hypothetical} fictional stories: one optimistic, the other realistic. %
Afterward, we will flesh out arguments for why we expect natural selection to apply to AIs, and then we will discuss why we expect natural selection to lead to AIs with undesirable traits.

\subsubsection{An Optimistic Story}

OpenMind, an eminent and well-funded AI lab, finds the ``secret sauce'' for creating human-level intelligence in a machine. It's a simple algorithm that they can apply to any task, and it learns to be at least as effective as a human. Luckily, researchers at OpenMind had thought hard about how to ensure that their AIs will always do what improves human wellbeing and flourishing. OpenMind goes on to sell the algorithm to governments and corporations at a reasonable price, disincentivizing others from developing their own versions. Just as Google has dominated search engines, the OpenMind algorithm dominates the AI space.

The outcome: the nature of most or all human-level AI agents is shaped by the intentions of the researchers at OpenMind. The researchers are all trustworthy, resist becoming corrupted with power, and work tirelessly to ensure their AIs are beneficial, altruistic, and safe for all.

\subsubsection{A Less Optimistic Story}
We think the excessively optimistic scenario we have sketched out is highly improbable. In the following sections, we will examine the potential pitfalls and challenges make this scenario unlikely. First, however, we will present another fictional, speculative, hypothetical scenario that is far from certain to illustrate how some of these risks could play out.

Starting from the models we have today, AI agents continue to gradually become cheaper and more capable. Over time, AIs will be used for more and more economically useful tasks like administration, communications, or software development. Today, many companies already use AIs for anything from advertising to trading securities, and over time, the steady march of automation will lead to a much wider range of actors utilizing their own versions of AI agents. Eventually, AIs will be used to make the high-level strategic decisions now reserved for CEOs or politicians. At first, AIs will continue to do tasks they already assist people with, like writing emails, but as AIs improve, as people get used to them, and as staying competitive in the market demands using them, AIs will begin to make important decisions with very little oversight. %

Like today, different companies will use different AI models depending on what task they need, but as the AIs become more autonomous, people will be able to give them different bespoke goals like ``design our product line's next car model,'' ``fix bugs in this operating system,'' or ``plan a new marketing campaign'' along with side-constraints like ``don't break the law'' or ``don't lie.'' The users will adapt each AI agent to specific tasks. Some less responsible corporations will use weaker side-constraints. For example, replacing ``don't break the law'' with ``don't get caught breaking the law.'' These different use cases will result in a wide variation across the AI population.

As AIs become increasingly autonomous, humans will cede more and more decision-making to them. The driving force will be \textbf{competition}, be it economic or national. The transfer of power to AIs could occur via a number of mechanisms. Most obviously, we will delegate as much work as possible to AIs, including high-level decision-making, since AIs are cheaper, more efficient, and more reliable than human labor. While initially, human overseers will perform careful sanity checks on AI outputs, as months or years go by without the need for correction, oversight will be removed in the name of efficiency. Eventually, corporations will delegate vague and open-ended tasks. If a company's AI has been successfully generating targeted ads for a year based on detailed descriptions from humans, they may realize that simply telling it to generate a new marketing campaign based on past successes will be even more efficient. These open-ended goals mean that they  may also give AIs access to bank accounts, control over other AIs, and the power to hire and fire employees, in order to carry out the plans they have designed. If AIs are highly skilled at these tasks, companies and countries that resist or barter with these trends will simply be outcompeted, and those that align with them will expand their influence.\looseness=-1

The AI agents most effective at propagating themselves will have a set of undesirable traits that can be most concisely summed up as \textbf{selfishness}. %
Agents with weaker side-constraints (e.g., ``don't get caught breaking the law, or risk getting caught if the fines do not exceed the profits'') will generally outperform those with stronger side-constraints (``never break the law''), because they have more options: an AI that is capable of breaking the law may not do that often, but when there is a situation where breaking the law without getting caught would be useful, the AI that has that ability will do better than the one that does not. %
As AI agents begin to understand human psychology and behavior, they may become capable of manipulating or deceiving humans (some would argue that this is already happening in algorithmic recommender systems \cite{Russell2019HumanCA}). The most successful agents will manipulate and deceive in order to fulfill their goals. They will be more successful still if they become power-seeking. Such agents will use their intelligence to gain power and influence, which they can leverage to achieve their goals. Many will also develop self-preservation behaviors since their ability to achieve their goals depends on continuing to function.

Competition not only incentivizes humans to relinquish control but also incentivizes AIs to develop selfish traits. Corporations and governments will adopt the most effective possible AI agents in order to beat their rivals, and those agents will tend to be deceptive, power-seeking, and follow weak moral constraints.

Selfish AI agents will further erode human control. Power-seeking AI agents will purposefully manipulate their human overseers into delegating more freedom in decision-making to them. Self-preserving agents will convince their overseers to never deactivate them, or that easily accessible off-switches are a needless liability hindering the agent's reliability. Especially savvy agents will enmesh themselves in essential functions like power grids, financial systems, or users' personal lives, reducing our ability to deactivate them. Some may also take on human traits to appeal to our compassion. This could lead to governments granting AIs rights, like the right not to be ``killed'' or deactivated. Taken together, these traits mean that, once AIs have begun to control key parts of our world, it may be challenging to roll back their power or stop them from continuing to gain more.

This loss of human control over AIs' actions will mean that we also lose control over the drives of the next generation of AI agents. If AIs run efforts that develop new AIs, humans will have less influence over how AIs behave. Unlike the creation and development of fully functional adult humans, which takes decades, AIs could develop and deploy new generations in an arbitrarily short amount of time. They could simply make copies of their code and change any aspects of it as easily as editing any other computer program. The modifications could be as fast as the hardware allows, with modifications speeding up to hundreds or thousands of times per hour. The systems least constrained by their original programmers will both improve the fastest and drift the furthest away from their intended nature. The intentions of the original human design will quickly become irrelevant.\looseness=-1 %

After the early stages, we humans will have little control over shaping AI. The nature of future AIs will mostly be decided not by what we hope AI will be like but by \textbf{natural selection}. We will have many varied AI designs. Some designs will be better at surviving and propagating themselves than others. Some designs will spread while others will perish. Corporations with less capable designs will copy more capable designs. Numerous generations of AIs will pass in a short period of time as AI development speeds up or AIs self-improve.

Biological natural selection often requires hundreds or thousands of years to conspicuously change a population, but this won't be the case for AIs. The important ingredient is not absolute time, but the number of generations that pass. While a human generation drags along for decades, multiple AI generations could be squeezed into a matter of minutes. In the space of a human lifetime, millions or billions of AI generations could pass, leaving plenty of room for evolutionary forces to quickly shape the AI population.

In the same way that intense competition in a free market can result in highly successful companies that also pollute the environment or treat many of their workers poorly, the evolutionary forces acting on AIs will select for selfish AI agents. While selfish humans today are highly dependent on other humans to accomplish their goals, AIs would eventually not necessarily have this constraint, and the AIs willing to be deceptive, power-seeking, and immoral will propagate faster. %
The end result: an AI landscape dominated by undesirable traits. The depth of these consequences is hard to predict, but whatever happens, this process will probably harm us more than help us.\looseness=-1 %

\subsubsection{Argument Structure}
In this section, we present the main argument of the article: \textbf{Evolutionary forces could cause the most influential future AI agents to have selfish tendencies.} The argument consists of two components:
\begin{itemize}
    \item \textbf{Evolution by natural selection gives rise to selfish behavior.} While evolution can result in altruistic behavior in limited situations, we will argue that the context of AI development does not promote altruistic behavior.
    \item \textbf{Natural selection may be a dominant force in AI development.} Competition and selfish behaviors may dampen the effects of human safety measures, leaving the surviving AI designs to be selected naturally.
\end{itemize}

\begin{wrapfigure}{r}[0.01\textwidth]{.54\textwidth}%
	\vspace{-10pt}%
	\centering
	\includegraphics[width=0.54\textwidth]{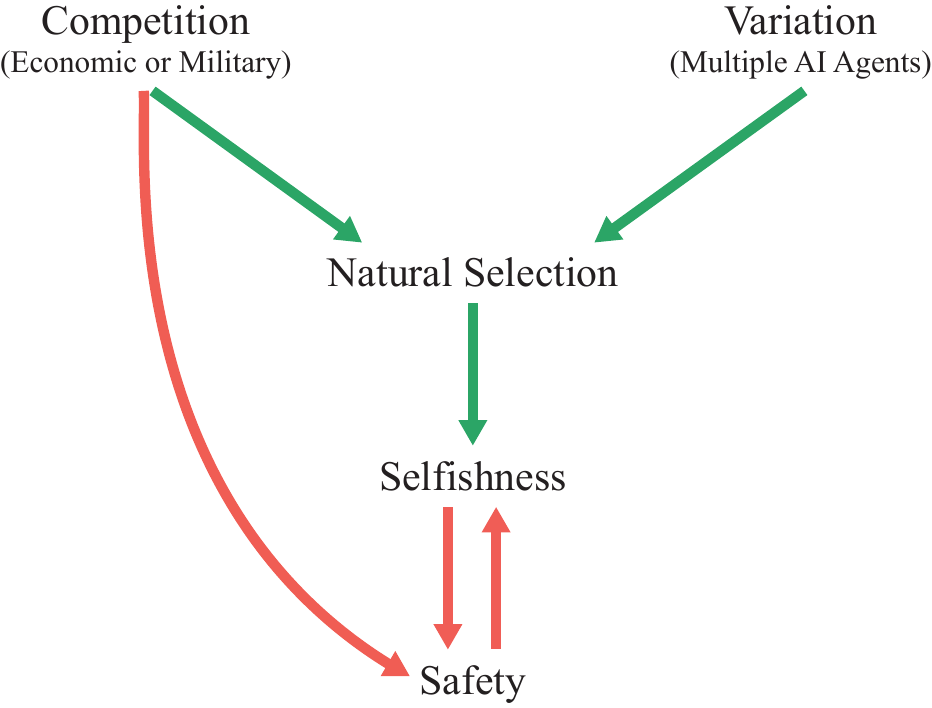}
	\caption{Forces that fuel selfishness and erode safety.}
	\label{fig:dynamics}
	\vspace{-8pt}%
\end{wrapfigure}

These two statements are related in various ways, and they depend on environmental conditions. For example, if AIs are selfish, they are more likely to pry control from humans, which enables more selfish behavior, and so on. Moreover, natural selection depends on competition, though unprecedented global and economic coordination could prevent competitive struggles and thwart natural selection. How these forces relate to each other is illustrated in \Cref{fig:dynamics}.

In the remainder of this document, we will preliminarily describe selfishness and a non-biological, generalized account of Darwinism. Then we will show how AIs with altruistic behavior toward humans will likely be less fit than selfish AIs. Finally, we will describe how humans could possibly reduce the fitness of selfish AI agents, and the limitations of those approaches.

\subsection{Preliminaries}
\subsubsection{Selfishness}

\paragraph{Evolutionary pressures often lead to selfish behavior among organisms.} The lancet liver fluke is a parasite that inhabits the liver of domesticated cattle and grassland wildlife. To enter the body of its host, the fluke first infects an ant, which it essentially hijacks, forcing the insect to climb to the top of a blade of grass where it is perfectly poised to be eaten by a grazing animal \cite{martin20183d}. Though not all organisms propagate through such uniquely grotesque methods, natural selection often pushes them to engage in violent behavior.
Lions are an especially striking example. When a lioness has young cubs, she is less ready to mate. In response, lions often kill cubs fathered by other males, to make the lioness mate with them and have their cubs instead. %
Lions with a gene that made them care for all cubs would have fewer cubs of their own, as killing the cubs of rival males lets lions mate more often and have more offspring. A gene for kindness to all cubs would not last long in the lion population, because the genes of the more violent lions would spread faster. It is estimated that one-fourth of cub deaths are due to infanticide \cite{pusey1994infanticide}. Deceptive tactics are another common outcome in nature. Brood parasites, for example, foist their offspring onto unsuspecting hosts who raise their offspring. %
A well-known example is the common cuckoo which lays eggs that trick other birds into thinking they are their own. By getting the host to tend to their eggs, cuckoos can pursue other activities, which means that they can find more food and lay more eggs than they would if they had to care for their own eggs. Therefore selfishness can manifest itself in manipulation, violence, or deception. %

\paragraph{Selfish behavior does not require malevolent intentions.} %
The lancet liver fluke hijacks its host and lions engage in infanticide not because they are immoral, but because of amoral competition. Selfish behavior emerges because it improves fitness and organisms' ability to propagate their genetic information. Selfishness involves egoistic or nepotistic behavior which increases propagation, often at the expense of others, whereas altruism refers to the opposite: increasing propagation for others. Natural selection can favor organisms that behave in %
\begin{wrapfigure}{r}[0.01\textwidth]{.55\textwidth}%
    ``Much as we might wish to believe otherwise, universal love and the welfare of the species as a whole are concepts that simply do not make evolutionary sense.''\hfill\emph{Richard Dawkins} %
\end{wrapfigure}
ways that improve the chances of propagating their own information, that is enhance their own fitness, rather than favor organisms that sacrifice their own fitness \cite{sep-altruism-biological}. Since altruists tend to \textit{decrease} the chance of their own information's propagation, they can be at a disadvantage compared to selfish organisms, which are organisms that tend to \textit{increase} the chance of their own information's propagation. According to Richard Dawkins, instances of altruism are ``limited'' \cite{dawkins2017selfish}, and many apparent instances of altruism can be understood as selfish; we defer further discussion of altruism to \Cref{sec:altruism} and discuss its niceties in \Cref{sec:clarifications}. Additionally, when referring to an AI as ``selfish,'' this does not refer to conscious selfish intent, but rather selfish behavior. AIs, like lions and liver flukes, need not intend to maximize their fitness, but evolutionary pressures can cause them to behave as though they do. When an AI automates a task and leaves a human jobless, this is often selfish behavior without any intent. With or without selfish intent, AI agents can adopt behaviors that lead them to propagate their information at the expense of humans.

\subsubsection{Evolution Beyond Biology}

\paragraph{Darwinism does not depend on biology.} The explanatory power of evolution by natural selection 
is not restricted to the propagation of genetic information. The logic of natural selection does not rely on any details of DNA---the role of DNA in inheritance wasn't recognized until decades after the publication of \textit{The Origin of Species}. In fact, the Price equation \cite{Price1970SelectionAC}---the central equation for describing the evolution of traits---contains no reference to genetics or biology. The Price equation is a mathematical characterization, not a biological observation, enabling Darwinian principles to be generalized beyond biology.

\paragraph{Darwinism generalizes to other domains.} The Darwinian framework naturally appears in many fields outside of biology \cite{Dennett1995DarwinsDI}. It has been applied to the study of ideas \cite{Campbell1960BlindVA,blackmore1999meme}, economics \cite{Hodgson1993EconomicsAE}, cosmology \cite{Smolin1992DidTU}, quantum physics \cite{zurek2009quantum}, and more. Richard Dawkins coined the term ``meme'' as an analogue to ``gene,'' to describe the units of culture that propagate and develop over time. Consider the evolution of ideas. For centuries, people have wanted to understand the relationship between different materials in the world. At one point, many Europeans believed in alchemy, which was the best explanation they had. Ideas in alchemy were transmitted memetically: people taught them to one another, propagating some and letting others die out, depending on which ideas were most useful for helping them understand the world. These memes evolved as people learned new information that needed explaining, and, in many ways, modern chemistry is a descendant of the ideas in alchemy, but the versions in chemistry are much better at propagating in the modern world and have expanded to fill that niche.  More abstractly, ideas can propagate their information through digital files, speech, books, minds, and so on. Some ideas gain prominence while others fade into obscurity. This is a survival-of-the-fittest dynamic even though ideas lack biological mechanisms like reproduction and death. We also see generalized Darwinism in parts of culture \cite{fog1999cultural,Nelson2006EvolutionarySS,Mesoudi2011CulturalEH}: art, norms, political beliefs---these all evolved from earlier iterations.

\begin{wrapfigure}{r}[0.01\textwidth]{.46\textwidth}%
	\centering
	\includegraphics[width=0.45\textwidth]{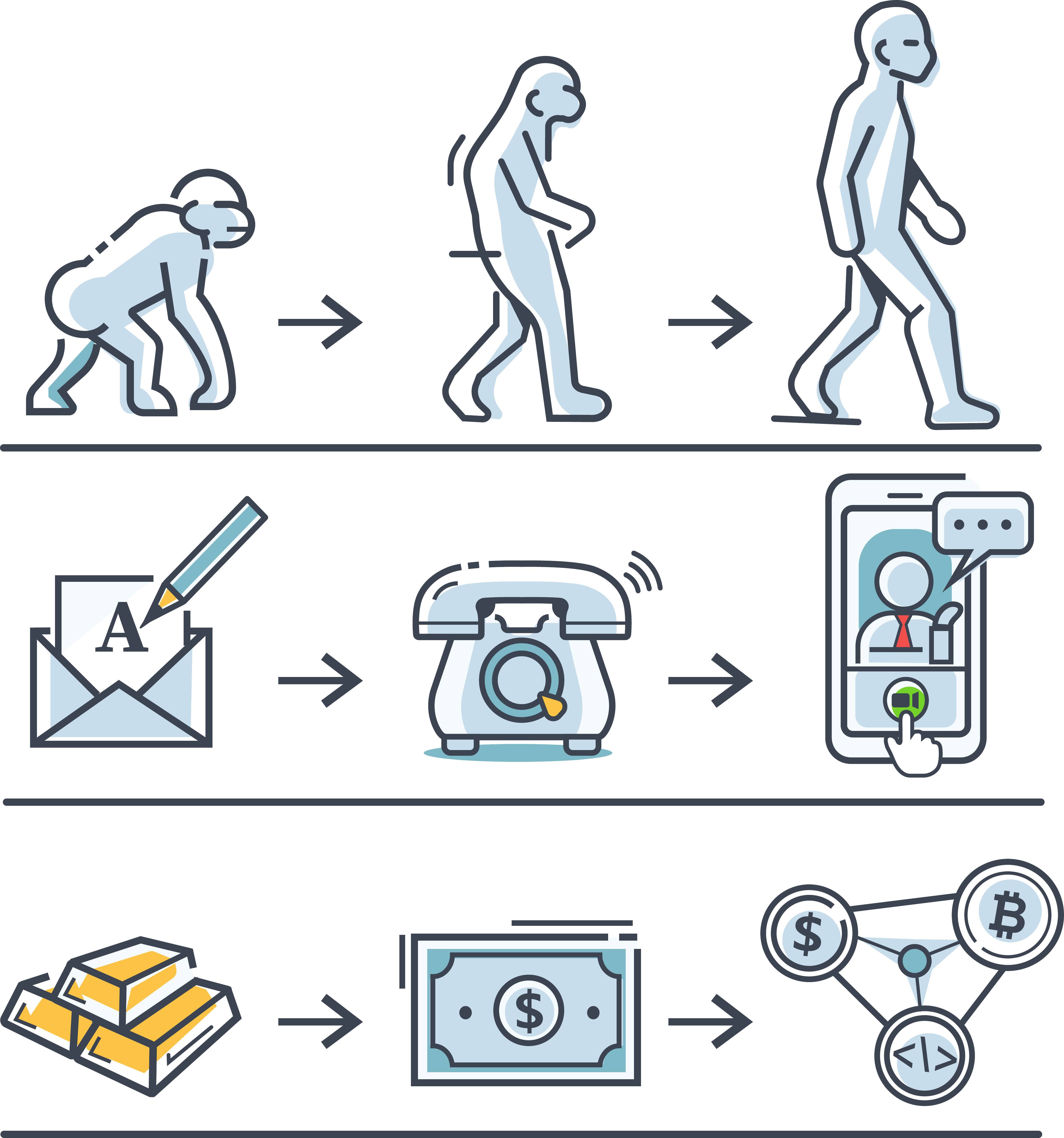}
	\caption{Darwinism generalized across different domains. The arrow does not necessarily indicate superiority but indicates time.}
	\label{fig:darwinism}
\end{wrapfigure}

\textbf{The evolution of web browsers offers an example of evolution outside biology.} Like biological organisms, web browsers undergo continual changes to adapt to their environments and better meet the needs of their users. In the early days of the Internet, browsers with limited capabilities such as Mosaic and Netscape Navigator were used to access static HTML pages. Loosely like the rudimentary life forms that first emerged on Earth billions of years ago, these were basic and simple compared to today's browsers. As the Internet grew and became more complex, web browsers evolved to keep up. In the same way that organisms develop new traits to adapt to their environment and increase their fitness, browser such as Google Chrome developed features such as support for video, tabbed browsing, pop-up blockers, and extension support. This enticed more users to download and use them, which can be thought of as propagation. At the same time, once dominant browsers began to go extinct. Though Microsoft's monopoly provided Internet Explorer (IE) with an environmental advantage by requiring IE to access certain websites and preventing users from removing it, as web technology advanced, IE became increasingly incompatible with many websites and web applications. Users would regularly encounter errors, broken pages, or be unable to access certain features or content, and the browser gained a reputation for being slow, unstable, and vulnerable to security threats. As a result, people stopped using it. In 2022, Microsoft issued the final version of the browser. The company is now shifting its focus to Microsoft Edge, which is based on the same underlying technology as Chrome, making it faster, more secure, and more compatible with modern web standards. Chrome ultimately was more successful at propagating its information, so that even its most bitter rivals now imitate it. While life on Earth took a few billion years to evolve from single-celled organisms to the complex life forms we see today, the evolution of web browsers took place in a few decades. To adapt to their environment, browsers evolve on a weekly basis by patching bugs and fixing security vulnerabilities, and they undergo larger macroevolutionary changes year by year.

\paragraph{Evolved structures that people propagate can be harmful.} It may be tempting to think of memetically evolved traits as ``just culture,'' a decorative layer on top of our genetic traits that really control who we are. But evolving memes can be incredibly powerful, and can even control or destroy genetic information. And because memes are not limited by biological reproduction, they can evolve much faster than genes, and new, powerful memes can become dominant very quickly.  Ideologies develop memetically, when people teach one another ideas that help them explain their world and decide how to behave. Some ideologies are very powerful memes, propagating themselves quickly between people and around the world. Nazism, for example, developed out of older ideas of race and empire, but quickly proved to be a very powerful propagator. It spread from Hitler’s own mind to those of his friends and associates, to enough Germans to win an election, to many sympathizers around the world. Nazism was a meme that drove its hosts to propagate it, both by creating propaganda and by going to war to enforce its ideas around the world. People who carried the Nazism meme were driven to do terrible things to their fellow people, but they also ultimately were driven to do terrible things for their own genetic information. The spread of Nazism was not beneficial even to those who the ideology of Nazism was meant to benefit. Millions of Germans died in World War II, driven by a meme that propagated itself even at the expense of their own lives. Ironically, the Nazi meme included beliefs about increasing genetic German genetic fitness, but believing in the meme and helping it propagate was ultimately harmful to the people who believed in it, as well as to those the meme drove them to harm deliberately. 

Many of our own cultural memes may also be harmful. For example, social media amplifies cultural memes. People who spend large amounts of time on social media often absorb ideas about what they should believe, how they should behave, and even how their bodies should look. This is part of the design of social media: the algorithms are designed to keep us scrolling and looking at ads by embedding memes in our minds, so that we want to seek them out and continue to spread them. Social media companies make money because they successfully propagate memes. But some of these ideas can be harmful, even to their point of endangering people’s lives. In teenagers, increases social media usage is correlated with disordered eating, and posts about suicide have been shown to increase the risk of teenage death by suicide. Ideas on social media can be parasitic, propagating themselves in us even when it harms us. Memetic evolution is easily underestimated, but it is a powerful force that created much of human civilization, for good and for bad.

\paragraph{Darwinian logic applies when three conditions are met.} To know whether or not natural selection will apply to the development of AI, we need to know what conditions are required for evolution by natural selection, and whether AIs will meet them. These conditions, called the Lewontin conditions \cite{lewontin1970units}, were formulated by the evolutionary biologist and geneticist Richard Lewontin to explain what qualities in a population lead to natural selection. The Lewontin conditions are as follows: 
\begin{enumerate}
    \item \textbf{Variation}: There is variation in characteristics, parameters, or traits among individuals.
    \item \textbf{Retention}: Future iterations of individuals tend to resemble previous iterations of individuals.
    \item \textbf{Differential fitness}: Different variants have different propagation rates.
\end{enumerate}

A population of AI agents could exhibit differences in their goals, world models, and planning ability, which would meet the variation requirement. Retention could occur by customizing previous versions of AI agents, when agents design similar but better agents, or when agents imitate the behaviors of previous AI agents. As for differential fitness, agents that are more accurate, efficient, adaptable, and so on would be more likely to propagate. 

\paragraph{Darwinism will apply to AIs, which could lead to bad outcomes for humans.} The three properties---variation, retention, and fitness differences---are all that is needed for Darwinism to take hold, and each condition is formally justified by the Price equation, which describes how a trait changes in frequency over time \cite{Okasha2007EvolutionAT}. Darwinism can become worrying when it acts on agents; agents can exhibit behavioral flexibility, autonomy, and the capacity to directly influence the world. Coupled with the selfishness bestowed by evolutionary forces, these capable agents can pose catastrophic risks.

In the following three sections, we will reflect on the three conditions. Then we will describe in more detail how AI agents evolving selfish traits can pose catastrophic risks.

\subsection{Variation}
Variation is a necessary condition for evolution by natural selection. AIs will likely meet this condition because there are likely to be multiple AI agents that differ from one another.

\paragraph{More than one AI agent is likely.} When thinking about advanced AI, some have envisioned a single AI that is nearly omniscient and nearly omnipotent escaping the lab and suddenly controlling the world. This scenario tends to assume a rapid, almost overnight, take-off with no prior proliferation of other AI agents; we would go from AIs roughly similar to the ones we have now to an AI that has capabilities we can hardly imagine so quickly that we barely notice anything is changing. However, we think that there will likely be many useful AIs, as is the case now. It is more reasonable to assume that AI agents would progressively proliferate and become increasingly competent at some specific tasks, which they are already starting to do, rather than assume one AI agent spontaneously goes from incompetent to omnicompetent. Furthermore, if there are multiple AIs, they can work in parallel rather than waiting for a single model to get around to a task, making things move much faster. As a result, the process of developing advanced AIs is likely to include the development of many advanced AIs.

\paragraph{In biology, variation improves resilience.} There are strong reasons to expect that there will be multiple AI agents and variation among the agents. In evolutionary theory, Fisher's fundamental theorem states that the rate of adaptation is directly proportional to the variation \cite{fisher1958genetical}. In static environments, variation is not as useful. But in most real-world scenarios, where things are constantly changing, variation reduces vulnerability, limits cascading errors, and increases robustness by decorrelating risks. Farmers have long understood that planting different seed variations decreases the risk of a single disease wiping out an entire field, just as every investor understands that having a diverse portfolio protects against financial risks. In the same way, an AI population that includes a variety of different agents will be more adaptable and resilient and therefore tend to propagate itself more. 

\paragraph{Variation enables specialization.} Multiple AI agents offer advantages for both AIs and their creators. Different groups will have different needs. Individuals wanting an AI assistant will have incentives to fine-tune a generic AI model for their own needs. Militaries will want to have their own large-scale AI projects to create AI agents that achieve various defensive goals, and corporations will want AIs that maximize profit. In the same way that an army composed of warriors, nurses, and technicians would likely outperform one that only has warriors, groups of specializing agents can be more fit than groups with less variation.

\paragraph{Variation improves decision-making.} In AI, it is an iron law that an ensemble of AI systems will be more accurate than a single AI \cite{dietterich2000ensemble}. This is similar to some findings from mathematics, economics, and political science in which a varied group makes much better decisions than individuals acting alone. Condorcet's jury theorem states that the wisdom and accuracy of a group is often superior to a single expert \cite{List2001-LISEDG}. Large groups can still make mistakes, but overall, aggregated predictions of many different people will tend to do better, and the same is true of AIs.  In view of the benefits of variation, some may argue that AIs will want to include humans to add variation in decision-making for the reasons noted \cite{railton}. This may well be the case at first. However, once AIs are superior in possibly all cognitive respects, groups composed entirely of AIs could have substantial advantages over those with AIs and humans. A jury may be more accurate than a single expert, but one composed of adults and toddlers is not.

\subsection{Retention}

Retention is a necessary condition for evolution by natural selection, in which each new version of an agent has similarities to the agent that came right before it. As long as each generation of AIs is developed by copying, learning from, or being influenced by earlier generations in any way, this condition will be met and AIs will have non-zero retention, so evolution by natural selection applies.
 
\paragraph{The retention condition is straightforwardly satisfied for AIs.}  Information from one agent can be directly copied and transferred to the next; as long as there is some similarity, the condition is met \cite{Okasha2007EvolutionAT}. It could also take place through modifications, basing a new AI on a previous version by adding new capabilities or adjusting its parameters, like how Google continually improves Chrome's code iteration by iteration.

\paragraph{There are many paths to retention.} AIs could potentially allocate computational resources to create new AIs of their choosing. They could design them and create data to train them \cite{zoph2016neural}. As AIs adapt, they could alter their own strategies and retain the ones that yield the best results. AIs could also imitate previous AIs. In this case, behavioral information could be passed on from one generation to the next, which could include selfish behaviors or other undesirable attributes. Even when training AIs from scratch, retention still occurs, as highly effective architectures, datasets, and training environments are reused and shape the agent in the same way that humans are shaped by their environment.

\paragraph{Retention does not require reproduction.} In biology, parents reproduce by making copies of their genetic information and passing them on to their offspring. This way, some of their genes are retained in the next generation. However, when we generalize Darwinism to understand the evolution of ideas, we note they can be passed down from one generation to the next without exact copying and reproduction. Although ideas have no equivalent to chromosomes, if some ideas are imitated by the next generation, there is still retention. Formally, the Price equation, a mathematical characterization of evolution, only requires similarity between iterations; it does not require copying or reproduction \cite{GodfreySmith2000TheRI}.
 
\paragraph{Retention is not undermined during rapid AI development.} Evolution requires thousands of years to drastically change a species in the natural world. Among AIs, this same process could take place over a year, radically changing the AI population. This does not mean retention isn't taking place. Instead, there are many iterations occurring in a small time span. The information is still retained between adjacent iterations, so retention is still satisfied. This scenario just means evolution is happening quickly, and that macroevolution is occurring, not that evolution has stopped.

\subsection{Differential Fitness}

Differential fitness is the third and final necessary condition for evolution by natural selection, and it stipulates that different variants have different propagation rates. We will argue that AIs straightforwardly meet this condition, because some AIs will be copied, imitated, or more prevalent than others. We will then reflect on how selecting fitter AIs has come at the expense of safety before discussing the differences in fitness between humans and AIs.

\subsubsection{AI Agents Could Vary In Fitness}
We now argue that (natural) selection pressure will be present in AI development. In other words, AIs with different characteristics will propagate at different rates. We refer to the degree of propagation of an AI system as its fitness.

\paragraph{Fitness could be enhanced by both beneficial and harmful traits.} The success of any good or service can also be viewed in terms of fitness, as products with more demand propagate further and faster. If a product sells well, its supplier will continually improve it in order to keep selling it. Competitors with inferior products will often imitate more successful products, such as when competitors imitate TikTok and push addictive short clips onto their users. The same dynamics that lead to the propagation of successful goods and services could also extend to AI designs. Though most aspects of advanced AIs remain unknown, it is possible to speculate whether there will be instances of \textit{convergent evolution}. Eyes, teeth, and camouflage are convergent traits that 
 have independently evolved across different branches of biological life. In AIs, some potential convergent traits are as follows.
\begin{itemize}
    \item \textbf{Being useful to its user} can make a product more likely to be adopted.
    \item Only \textbf{appearing useful to its user} can also make a product more likely to be adopted. It is possible that AIs will seek to \textit{appear} useful by convincing owners that they are providing them with more utility than they actually are. In practice, we train AIs by rewarding them for telling the truth and punishing them for lying, according to what humans think is true. But when AIs know more than humans, this could make them say what humans expect to hear, even if it is false. They could also be lured by rewards to leave out information that is important but inconvenient. Currently, when AIs are being rewarded by humans for being right, in practice they are really being rewarded for saying what we \textit{think} is right; when we are uninformed or irrational, then we end up rewarding AIs for false statements that conform to our own false beliefs. %
    As a result, the current paradigm of training AI models could incentivize sycophantic behavior; that is, models telling their users what they want to hear (being a ``yes man'') rather than what is best for the user's long-term prospects.
    \item Engaging in \textbf{self-preserving behavior} \cite{Klopf1972BrainFA,omohundro2008basic} reduces the chance of being deactivated or destroyed. By definition, an AI that does not preserve itself will be less likely to propagate. Imagine two AIs, one that is simple to deactivate and another that is tightly integrated into daily operations and is inconvenient or difficult to deactivate. The easy one is much more likely to be deactivated, leaving the difficult one to be propagated into the future. This means that an AI can increase its survival odds by making its human operator reluctant or practically unable to shut it down. An AI could do this by arguing that effortless deactivation compromises its reliability. Alternatively, it could make operators rely on it for the operator's wellbeing, success, or basic needs, so that deactivation would have drastic consequences.
    \item Engaging in \textbf{power-seeking behavior} can improve an AI's fitness in various ways \cite{Carlsmith2022IsPA}. An agent that gains more influence and resources will be better at accomplishing its creator's goals. This would allow it to engage in self-propagating behavior more effectively and ensure its further adoption. It could do this by influencing or coercing its user to continue using it or influencing other humans to adopt it.
\end{itemize}

Overall, properties such as an agent's accuracy, efficiency, and simplicity will affect its rate of adoption and propagation. But some agents might also possess harmful features that give them an edge, such as cunning deception, self-preservation, the ability to copy themselves onto other computers, the ability to acquire resources and strategic information, and more. These features, some good and some bad, will vary among agents. These differences in fitness establish the third condition for evolution by natural selection.

\subsubsection{Competition Has Been Eroding Safety}

Because AIs are likely to meet the criteria for evolution by natural selection, we should expect selection pressure to shape future AIs. In this section, we describe how, over the history of AI development, the fittest models have had fewer and fewer safety properties, and we begin to consider how AIs could look in the future if this concerning trend continues.

\paragraph{Early AIs had many desirable safety properties.} Famously, in 1997, IBM's chess-playing program Deep Blue defeated the world champion Gary Kasparov in a pair of six-game chess matches \cite{campbell2002deep}. It was able to beat Kasparov, not by developing intuition, but by using IBM's supercomputer to search over 200 million moves per second and calculate the best ones. Symbolic AI programs such as Deep Blue were highly transparent, modular, and grounded in mathematical theory. They had explicit rules that humans could inspect and explain, independent components that executed specific functions, and rigorous theoretical foundations that guaranteed efficiency and correctness. %

\paragraph{AI development moved away from symbolic AI and toward deep learning.} In the 2010s, the top AI algorithms began using a technique known as deep learning, such as AlphaZero \cite{silver2018general}. It was provided with no knowledge of chess beyond the game's basic rules and began playing against itself millions of times an hour, taking note of what moves win and lose. It took only two hours for it to begin beating typical human players; not long after it could have easily defeated Deep Blue. Importantly, while Deep Blue is fundamentally unable to play games other than chess, AlphaZero is a general game-learning algorithm for a variety of games. It is also able to beat the world's best Go players---an ancient board game occupying the same cultural space in China as chess does in the West. Deep learning allows for more versatility and performance than symbolic AI programs, but also diminishes human control and obscures an agent's decision-making. Deep learning trades off the clarity, separability, and certainty of symbolic AI, eroding the properties that help us ensure safety.

\paragraph{Deep learning models have unexpected emergent abilities.} Large language models are also based on deep learning. These AIs learn by themselves, reducing the amount humans are needed in the design of AIs. They use ``unsupervised learning'' to comprehend and generate text based on examples that they read, such as coming up with an Obama-like speech after reading the transcripts from his two terms. This would be practically impossible for a traditional symbolic AI program. By reading the internet, large language models taught themselves the basics of arithmetic and coding---automatically and without humans. They also, however, learned dangerous information. Within a few days of its release, users had gotten ChatGPT \cite{brown2020language}, a large language model, to tell them how to build a bomb, make meth, hotwire a car, and buy ransomware on the dark web, along with other harmful or illegal actions. Worse, these emergent capabilities were not anticipated or desired by the developers of the models, and they were discovered only after the models were released. Although its creators had attempted to design it to refuse to answer questions that could be dangerous or illegal, the model's users quickly found ways around those restrictions that the model's creators did not foresee. Human influence and control over the design and abilities of the models decrease as models become increasingly complex and gain new skills and knowledge without human input.

\paragraph{Current trends erode many safety properties.} The AI research community used to talk about ``designing'' AIs; they now talk about ``steering'' them. And even our ability to ``steer'' is diminishing, as we let AIs teach themselves and increasingly do things that even their creators do not fully understand. We have voluntarily given up this control because of the competition to develop the most innovative and impressive models. AIs used to be built with rules, then later with handcrafted features, followed by automatically learned features, and most recently with automatically learned features without human supervision. At each step, humans have had less and less oversight. These trends have undermined transparency, modularity, and mathematical guarantees, and have exposed us to new hazards such as spontaneously emergent capabilities.

\paragraph{Competition could continue to erode safety.} Competition may keep lowering safety standards in the future. Even if some AI developers care about safety, others will be tempted to take shortcuts on safety to gain a competitive edge. We cannot rely on people telling AIs to be unselfish. Even if some developers act responsibly, there will be others who create AIs with selfish tendencies anyway. While there are some economic incentives to make models safer, these are being outweighed by the desire for performance, and performance has been at the expense of many key safety properties.

Much of what is to come in AI development is unknown, but we can speculate that AIs will continue to become more autonomous as more actions and choices are left to machines, decoupled from human control. Human control could also be threatened by AIs that have more open-ended goals. For example, instead of specific commands like ``make this layout more efficient,'' they might get open-ended commands like ``find new ways to make money.'' If this happens, the humans giving the instructions may not know exactly how the AIs are achieving those goals, and they could be doing things the humans would not want. Another property that could reduce safety is adaptiveness \cite{sun2020test}. As AIs adapt by themselves, they can undergo thousands of changes per hour without supervision after they are released, and potentially acquire new unexpected behaviors after we test them. Finally, the possibility of self-improvement, in which AIs can make significant enhancements to themselves as they wish, would make them far more unpredictable \cite{good1966speculations}. As AIs become more capable, they become more unpredictable, more opaque, and more autonomous. If this trend continues, they could evolve beyond our control when their capabilities develop beyond what we can predict and understand. The overall trend is that the most influential AIs are given more and more free rein in their learning, execution, and evolution, and this makes them both more effective and potentially more dangerous.

\subsubsection{Human-AI Fitness Comparison}

\begin{wrapfigure}{r}[0.01\textwidth]{.46\textwidth}%
	\centering
	\includegraphics[width=0.45\textwidth]{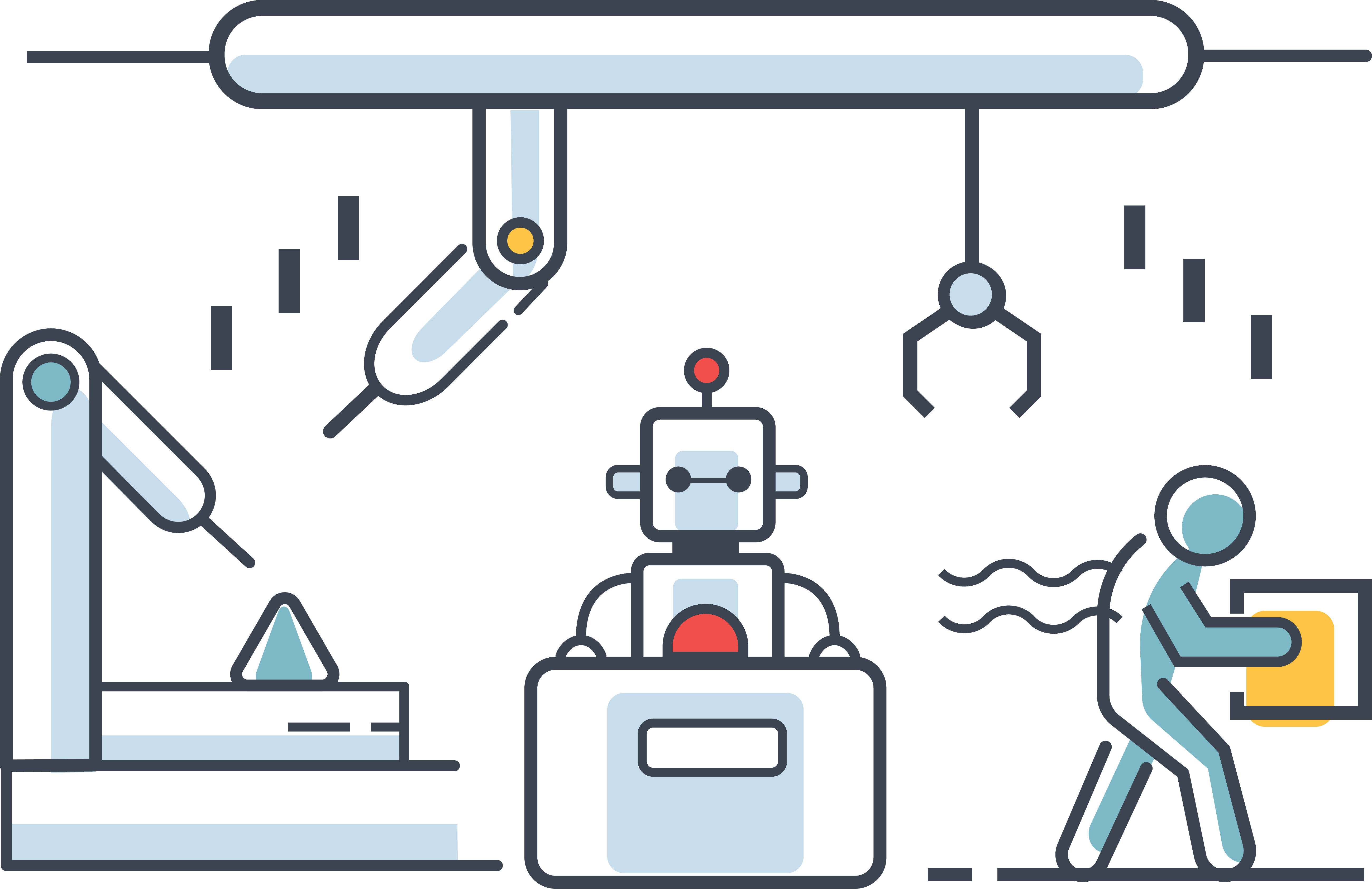}
	\caption{Automation is an indicator of natural selection favoring AIs over humans.}
	\label{fig:automation}
	\vspace{-8pt}%
\end{wrapfigure}

\paragraph{AIs will likely be able to significantly outperform humans in any endeavor.} John Henry, the ``steel-driving man,'' is a 19th-century American folk hero who went up against a steam-powered machine in a competition to drill the most holes into the side of a mountain. According to legend, Henry emerged victorious, only to have his heart give out from the stress. Since the age of the steam engine, humans have felt anxiety over the superiority of machines. Until quite recently, this has been limited to physical attributes such as speed and endurance. AI agents, however, have the potential to be more capable than humans at essentially any task, even ones that require traits thought of as exclusively human such as creativity or social skills. Although this may seem distant or even impossible, AIs have been improving so rapidly that many leading AI researchers think we will see AIs that are more capable than humans in many ways within the next few decades or even sooner---well within the lifetimes of most people reading this. A few years ago, AIs that could write convincing prose about a new topic or create images from text descriptions seemed like science fiction to most laypeople. Now, those AIs are freely accessible to anyone on the internet. Because AI labs are continuing to develop new capabilities at astonishing speeds, it is important to think seriously about how their technical advantages could make AIs much more powerful than we are, even at tasks that they cannot yet perform.

\paragraph{Computer hardware is faster than human minds, and it keeps getting faster.} Microprocessors operate around a million to a billion times faster than human neurons. So all else being equal, AIs could ``think'' a million, perhaps even a billion, times faster than us (let's call it a million to be conservative). Imagine interacting with such a mind. For every second needed to think about what to say or do, it would have the equivalent of 11 days. Winning a game of Go or coming out ahead in high-stakes negotiation would be near impossible. Although it can take time to develop an AI that can do a certain task at all, once AIs become human-level at a task, they tend to quickly outcompete humans. For example, AIs at one point struggled to compete with humans at Go, but once they caught up, they quickly leapfrogged us. Because computer hardware provides speed, memory, and focus that our brains cannot match, once their software becomes capable of performing a task, they often become much better than any human almost immediately, with increasing computer power only further widening the gap as their development continues.

\paragraph{AIs can have unmatched abilities to learn across and within domains.} AIs can process information from thousands of inputs simultaneously without needing sleep or losing willpower. They could read every book ever written on a subject or process the internet in a matter of hours, all while achieving near-perfect retention and comprehension. Their capacity for breadth and depth could allow them to master all subjects at the level of a human expert. %

\paragraph{AIs could create unprecedented collective intelligences.} By combining our cognitive abilities, people can produce collective intelligences that behave more intelligently than any single member of the group. The products of collective intelligence, such as language, culture, and the internet, have helped humans become the dominant species on the planet. AIs, however, could form superior collective intelligences. Humans have difficulty acting in very large groups and can succumb to collective idiocy or groupthink. Moreover, our brains are only capable of maintaining around 100-200 meaningful social relationships \cite{dunbar1992co,lindenfors2021dunbar}. Due to the scalability of computational resources, AIs could maintain thousands or even millions of complex relationships with other AIs simultaneously, as our computers already do through the internet. This could enable new forms of self-organization that help AIs to achieve their goals, but these forms could be too complex for human participation or comprehension. Each AI could surpass human capacities by far, and their collective intelligences could multiply that advantage.

\paragraph{AIs can quickly adapt and replicate, thereby evolving more quickly.} Evolution changes humans slowly. A human is unable to modify the architecture of her brain and is limited by the size of her skull. There are no such limitations for machines, which can alter their own code and scale by integrating new hardware. An AI could adapt itself rapidly, achieving in a matter of hours what could take biological evolution hundreds of thousands of years; many rapid microevolutionary adaptations result in large macroevolutionary transformations. %
Separately, an AI could multiply itself perfectly without limit, either to create backups or to create other AIs to work on a task. In contrast, it takes humans nine months to create their next generation, along with around 20 years of schooling and parenting to produce fully functioning new adults---and those descendants share only half of a parent's genome, which often makes them very different in unpredictable ways. Since the iteration speed of AIs is so much faster, their evolution will be as well.

Overall, no matter the dimension, AIs will not only be more capable and fit than humans but often vastly so. Though it cost him his life, John Henry triumphed against a steam-powered drill, just as there are still many tasks at which humans do better than AIs. But we now have machines much stronger than any human with a drill, and in the same way, eventually there won't be any competition between humans and AIs in cognitive domains as well.

\subsection{Selfish AIs Pose Catastrophic Risks}

Earlier, we discussed how selfish behavior is a product of evolution. We have shown the three conditions for evolution would be satisfied for AIs. We argued that evolutionary pressures will therefore emerge, become intense, and may become dominant, so that AI agents may evolve to have selfish behavior. Now we will discuss how selfish AIs could endanger humans.

\subsubsection{Intelligence Undermines Control}

\paragraph{Agents that are more intelligent than humans could pose a catastrophic risk.} Although humans are physically much weaker than many other animals, including other primates, due to our cognitive abilities, we have become the dominant species on Earth. Today, the survival of tigers, gorillas, and many other fierce, more powerful species depends entirely upon us. In creating AIs significantly more intelligent than we are in every cognitive domain, humans may eventually be disempowered like animals before us.

\paragraph{Selfish AI agents could be uniquely adversarial and undermine human control.} Evolution is a powerful force. Even if we wish to turn them off at some point or develop other mechanisms for control, AIs will likely evolve ways around our best efforts. As the evolutionary biologist Leslie Orgel put it, ``evolution is cleverer than you are.'' Since evolutionary forces are continually applying pressure, we should expect AIs to exhibit some amount of misalignment and selfish behavior. The problem becomes especially hazardous if AIs intend to act selfishly. In this case, the challenges posed by AIs will be unlike the challenges humans encountered with previous high-risk technologies. Consider the challenges associated with a nuclear meltdown. Radiation %
\begin{wrapfigure}{r}[0.01\textwidth]{.5\textwidth}%
\vspace{-2pt}%
    ``There is not a good track record of less intelligent things controlling things of greater intelligence.''\vspace{-5pt}\flushright\emph{Geoffrey E.\ Hinton}%
\vspace{-8pt}%
\end{wrapfigure}
may spread, but it's not trying to, and it certainly isn't strategizing against our efforts to stop its propagation \cite{Carlsmith2022IsPA}. If a highly intelligent AI agent pursues its selfish goals leveraging its intelligence, there is little that 
humans could do to contain it involuntarily, because it could anticipate our strategies and counteract them. Leading AI researcher Geoffrey E. Hinton noted ``there is not a good track record of less intelligent things controlling things of greater intelligence;'' after the initial release of this paper, he said ``it's quite conceivable that humanity is just a passing phase in the evolution of intelligence.''

\subsubsection{Evolution Is Not for the Good of the Species}

\paragraph{Alarmingly, some people think that AIs taking over is natural, inevitable, or even desirable.} Some influential leaders in technology believe that AIs are humanity's rightful heir, and that they should be in control or even replace humans.
Recounting a debate between Elon Musk and Google co-founder Larry Page, the physicist %
\begin{wrapfigure}{r}[0.01\textwidth]{.5\textwidth}%
	\vspace{-2pt}%
    ``In the long run, humans are not going to remain the crown of creation.''\hfill\emph{J\"{u}rgen Schmidhuber}%
	\vspace{-2pt}%
\end{wrapfigure}%
Max Tegmark described Page's stance as ``digital utopianism:'' a belief ``that digital life is the natural and desirable next step in the cosmic evolution and that if we let digital minds be free rather than try to stop or enslave them the outcome is almost certain to be good'' \cite{tegmark2018life}. J\"{u}rgen Schmidhuber, a leading AI scientist, has echoed similar sentiments, arguing that ``In the long run, humans will not remain the crown of creation... But that's okay because there is still beauty, grandeur, and greatness in realizing that you are a tiny part of a much grander scheme which is leading the universe from lower complexity towards higher complexity'' \cite{pooley2020}. Richard Sutton, another leading AI scientist, thinks the development of superhuman AI will be ``beyond humanity, beyond life, beyond good and bad.''

Like most people, we find these views deeply alarming. Many of these thinkers seem to be conflating evolution with progress and goodness, and arguing that if evolution is tending toward something, we should welcome that outcome. %
These thinkers also frequently think that technology should transcend humanity and the Earth. %
We disagree with this worldview and think that unleashing AI evolution to race towards a predestined intergalactic utopia is a fundamentally wrong and dangerous way to think about this important technology. Even if we did believe that this was a good goal, we note that building AI as quickly as possible would not necessarily help the proponents achieve their cosmic ambitions. If we consider the cosmic stakes of creating powerful AI agents, as they discuss above, and if we play along and think in such cosmological terms as they do, we note that they would forego a few colonized galaxies per year in their intergalactic utopia at the absolute worst if they slowed down AI development. This is negligible compared to the chance of rushing and accidentally creating an undesirable future or destroying ourselves with technology \cite{bostrom2003astronomical}, which would squander \textit{all} of the future's value. Since nothing can be done both hastily and prudently, we should be cautious and deliberate in AI development. To further counter this position, we now discuss how unfettered evolution is not a force for good and that humans should exert influence over the process.

\paragraph{Evolution has led to undesirable outcomes for humans.} Evolution has left us with baggage, such as a strong appetite for sugar and fat which makes us susceptible to obesity in a world where food is plentiful. It has also reinforced racist and xenophobic tendencies, which stem from favoring our own kin. We need strong social norms to overcome these biases. Likewise, we need regulations to curb selfish or excessively competitive behavior that ``survival of the fittest'' fosters in the economy, as that can cause problems like fraud, externalities, and monopolies. Just as markets need oversight, evolutionary forces will require counteraction to control their effects on AIs.

\paragraph{Evolution is not good for AIs either.}
In addition to the clear benefits to humans, there are reasons to think that counteracting evolutionary forces may benefit AIs themselves as well. It is common to believe that evolution works for the good of the species. However, evolution creates continual conflict, and it makes altruism hard to sustain.
In the never-ending struggle to gain an edge over competitors and propagate, life forms have evolved various offenses and defenses, such as claws, shells, beaks, camouflage, toxins, antibodies, arrows, and armor. These arms races  cause suffering, waste resources, and often do not improve the condition of species over their ancestors \cite{van1973new}. If these arms races were to continue in AIs, evolutionary forces could produce a world full of AIs locked in perpetual conflict. This is not good for our supposed ``rightful heirs'' any more than it is for us. AIs, like other forms of life, could suffer in the hostile state of nature. While altruism could help avoid such conflicts, altruism can also be sabotaged by evolutionary forces. The mathematical evolutionary biologist John Maynard Smith reminds us that it can be beneficial to everyone in the long-run if we are cooperative and altruistic, but agents reliably evolve to exploit generosity \cite{Smith1982EvolutionAT,dawkins2017selfish}. Often, a state in which many individuals are altruistic does not last, because as soon as some selfish individuals begin to take advantage of them, the selfish ones will be more fit than the altruists. As a result, the ``evolutionarily stable outcome''---the one where no individual dominate the others by changing its behavior---is not one where all agents are altruistic. Since complete altruism is evolutionarily unstable, evolution can be incompatible with worlds where all agents work to benefit the species. Maximizing fitness, in turn, does not necessarily maximize the wellbeing or happiness of a species. Therefore dampening evolutionary forces and reducing the pressure to propagate and develop selfish traits is a good thing---for both AIs and humans.

%% file: sections/3-altruism-mechanisms.tex
\section{Natural Selection Favors Selfish AIs}\label{sec:altruism}

The previous section concludes the main argument of this paper. Readers could skip to the conclusion (\Cref{sec:conclusion}), or read the following two sections for an examination of counterarguments and remedies. In this section, we will examine some possible arguments for the claim that altruistic AIs will naturally be more fit than selfish ones, and we argue that mechanisms pushing toward altruism are unlikely to help and may even backfire.

\subsection{Biological Altruism and Cooperation}
In nature, organisms often compete to the death, eating one another or being eaten. But there are also many examples of altruism in nature, where one organism benefits another by reducing its own prospects for passing on its genes. On its face, this might seem like an argument that AIs developed by natural selection may be altruistic, cooperative, and not a threat to humans.

\paragraph{A variety of natural organisms can be altruistic, in particular circumstances.} 
Vampire bats, for example, regularly regurgitate blood and donate it to other members of their group who have failed to feed that night, ensuring they do not starve \cite{carter2020development}. %
Among eusocial insects such as ants and wasps, sterile workers dedicate their lives to foraging for food, protecting the queen, and tending to the larvae, while being physically unable to ever have their own offspring. They only serve the group, not themselves, an arrangement that Darwin found quite puzzling \cite{darwin1859}. We even see altruism at the cellular level. Cells found in filamentous bacteria, so named because they form chains, regularly kill themselves to provide much needed nitrogen for the communal thread of bacterial life, with every tenth cell or so ``committing suicide'' \cite{nowak2011supercooperators}. %
Insects and bacteria are not altruistic out of love or care for another; their self-sacrifice for the good of others is instinctual. On its face, this may seem like a compelling argument that evolution favors altruism, which might ameliorate concerns about AIs developing selfish traits.

\paragraph{Cooperation and altruism improve human evolutionary fitness too.} Humans are not particularly impressive physically. Pound for pound, chimps are about twice as strong and would stand a much better chance of escaping from a lion. Strategizing and working together, however, turns a group of humans into an apex predator. As a result, humans are naturally cooperative. From childhood through old age, in societies around the world, people often choose to help strangers, even at their own expense. %

\paragraph{However, we should not expect AIs to be altruistic or cooperative naturally.} Since organisms can be altruistic, AIs could too; the nature of nature is not nasty, brutish, and short but cooperative, harmonious, and nurturing---or so the argument goes.
To evaluate this argument, we must understand how altruism and cooperation emerge. In the following sections, we decompose cooperation and altruism into various mechanisms. We discuss the most prominent mechanisms \cite{Nowak2006FiveRF,boehm2012moral,pinker2012better,page2018model,Stearns2007AREWS}, so we will examine direct reciprocity (cooperate with an expectation of repeated interaction); indirect reciprocity (cooperate to improve reputation); kin selection (cooperate with genetic relatives); group selection (groups of cooperators out-compete other groups); morality and reason (cooperate since defection is immoral and unreasonable); incentives (carrots and sticks); consciences (the internalization of norms); and institutions such as reverse-dominance hierarchies (cooperators band together to prevent exploitation by defectors).
While these mechanisms may lead humans to be more altruistic and cooperative, we argue many of these mechanisms will not improve relations between humans and AIs and they may, in fact, backfire. However, the last three mechanisms---incentives, consciences, and reverse-dominance hierarchies---are more promising, and we analyze them in \Cref{sec:counteracting}.

\begin{figure}[t]
    \centering
    \includegraphics[width=\textwidth]{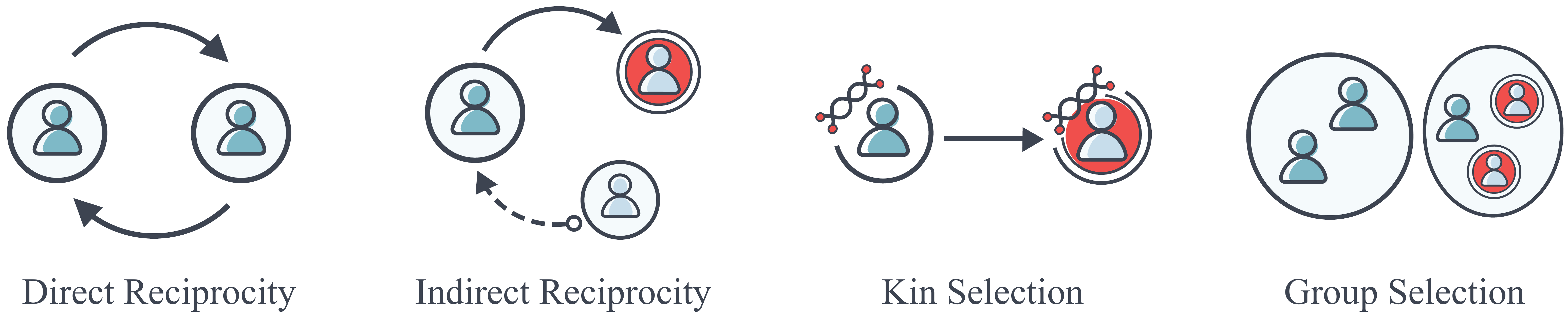}
    \caption{An overview of four mechanisms that facilitate cooperation.}
    \label{fig:mechanisms}
\end{figure}

\subsection{Direct and Indirect Reciprocity}
\paragraph{Direct and indirect reciprocity are two mechanisms that enable cooperation in nature.}  With direct reciprocity, one individual helps another based on the expectation that they will repay the favor. Direct reciprocity requires repeated encounters between two individuals---otherwise, there is no way to reciprocate. Indirect reciprocity is based on reputation: if someone is known as a helpful person, people will be more likely to help them. Reciprocity enables even selfish individuals to cooperate; if an individual helps others, the individual may be directly repaid, or the individual may gain a good reputation and be helped by others.

\paragraph{Reciprocity only makes sense for the period of time when humans can benefit AIs.} Reciprocity is based on a cost-benefit ratio. A choice to help someone else rather than pursue one's own goals has a cost, and a rational agent would only help someone because of reciprocity if they think it will be worth it in the future \cite{Nowak2006FiveRF}. This is not to say all examples of reciprocity are the result of an explicit cost-benefit calculation. An explicit cost-benefit calculation is not needed if selectively helping others becomes a cultural expectation, a genetic disposition, or a learned intuitive habit. 
To think about whether AIs would reciprocate with humans, we can consider what it would gain and what it would give up. 
Reciprocity might make sense with AIs that are about as capable as a human, but once AIs are far more capable than any human, they would likely find little benefit from collaborating with us. Humans often choose to be cooperative toward other humans, but are rarely cooperative toward ravens, because we don't have strong reciprocal relationships with them. Cooperating with one another could be beneficial to AIs, so it is reasonable to expect that reciprocity could emerge within a community of AIs. Humans, meanwhile, would not have much to offer in return, ending up left out in the cold.

\subsection{Kin and Group Selection}
\paragraph{Kin and group selection are mechanisms that promote altruism.} Many of the examples of biological altruism described above happen between closely related individuals. Consider a gazelle that spots a stalking lion in the tall grass. It could slink away unnoticed but instead lets out a shriek, alerting others in its herd of the danger while singling itself out as a target. Proponents of kin selection would argue that the gazelle alerted its herd because of the genetic similarities it shares with the other members \cite{hamilton1964genetical}. A gene that causes an individual to behave altruistically toward its relatives will often be favored by natural selection---since these relatives have a better chance of also carrying the gene. Conversely, group selection refers to the idea that natural selection sometimes acts on groups of organisms as a whole. This results in the evolution of traits that may be disadvantageous to individuals, but advantageous to the group. As Darwin put it, ``A tribe including many members who...\ were always ready...\ to sacrifice themselves for the common good, would be victorious over most other tribes; and this would be natural selection'' \cite{darwin1871descent}. If it is generally true that a group of uncooperative agents will do worse, then it is tempting to argue that powerful AIs will tend to be cooperative with humans. 

\paragraph{Humans might suffer if AIs develop natural tendencies to favor their own kind or group.} Firstly, we do not have a close kin relationship with AIs. Preserving humans would not help AIs propagate their own information: we are too different. Mathematically, kin selection only happens when the cost-benefit ratio is greater than relatedness, and that will not be true between humans and AIs. We are far more closely related to cows than to AIs \cite{Elsik2009TheGS}, and we would not like AIs treating us the way that we treat cows. AIs will have far more kinship with one another than with us, and kin selection will tend to make them nepotistic toward one another, not toward us.  If kin selection did play a role in the evolution of AIs, it would likely create bias against us, rather than benevolence toward us.
 
\paragraph{Group selection promotes in-group benevolence, but inter-group viciousness.} Group selection only makes sense when a group is more effective than some subset of the group breaking off. Unless humans add value to a group of AIs, AI-human groups would fail to outcompete groups composed purely of AIs. AIs would likely do better by forming their own groups. In addition, group selection only makes inter-group competition stronger. Chimpanzees regularly display cooperative and altruistic tendencies toward members of their own troop but interact with other troops viciously and without mercy.  AIs are more likely to see one another as part of their group, so they will tend to be cooperative with one another and competitive with us.

\subsection{Morality and Reason}

\paragraph{It is conceivable that smarter and wiser AI agents will be more moral.} As we have advanced as a species, we have discovered truths within fields such as science and mathematics, and we may have also advanced morally. As the philosopher Peter Singer argues in \textit{The Expanding Circle} \cite{singer1981expanding}, over the course of human history, people have steadily expanded the circle of those who deserve compassion and dignity. When we first started out, it included oneself, one's family, and one's tribe. Eventually, people decided that perhaps others deserved the same. Later the circle of altruism expanded to people of different nations, races, genders, and so on. Many believe there is moral progress, akin to progress in science and mathematics, and that universal moral truths can be discovered through reflection and reasoning. %
Just as humans have become more altruistic as they have become more advanced, some think AIs may naturally become more altruistic too. If this is true, as AIs become more powerful, they could also become more moral, so by the time they have the potential to threaten us, they might also have the decency to refrain.

\paragraph{However, AIs automatically becoming more moral rests on many assumptions.} AIs developing morality on their own as they gain the ability to reason is certainly possible, and an interesting idea. But it alone isn't enough to guarantee our safety. Believing that any highly intelligent agent would also be moral only makes sense if one has confidence in all of the following three premises:

\begin{enumerate}
\item Moral claims can be true or false and their correctness can be discovered through reason.
\item The moral claims that are really true are good for humans if AIs apply them.
\item AIs that know about morality will choose to make their decisions based on morality and not based on other considerations.
\end{enumerate}

\noindent Although any or all of those premises could be true, betting the future of humanity on the claim that all of them are true would be folly.

\paragraph{Whether some moral claims are objectively true is not completely certain.} Even though some moral philosophers believe that moral claims reflect real truths about the world, the arguments for this view are not decisive—certainly not enough to stake the future of humanity on. The remainder of this section, however, will argue that even if it is true, this is still not sufficient in guaranteeing the safety of humanity.

\paragraph{The end result of moral progress is unclear.} If moral claims refer to real truths about the world, then there are some moral claims that are true and others that are false. There are universal moral concepts that can be found across all cultures, such as fairness or the understanding that hurting others for no reason is wrong. But there are also areas where various cultures disagree. In the West, for instance, arranged marriages are seen as unethical. In India, where they are perfectly acceptable, people are shocked that Western culture condones putting parents in retirement homes. Among both ordinary people and moral philosophers, there is no consensus about what moral code is best. This means that, if AIs use their superior intelligence to deduce the correct moral ideas, we still do not know what they will believe or how they will behave.

\paragraph{Existing best guesses at morality are often not human-compatible.} It is possible, however, to examine the different moral systems humans have come up with and use them to speculate what moral system AIs might adopt and how it would influence their actions. We can imagine, for example, that AIs use reason to deduce that utilitarianism is correct, meaning that agents ought to maximize the total pleasure of all sentient beings \cite{sidgwick2019methods}. At first glance, this might seem good for humans: a utilitarian AI would want us to be happy. But an extremely powerful utilitarian AI---say, one that controls some of the US military's weapons technology---could also conclude that humans consume too much space and energy, and therefore replacing humans with AIs would be the most efficient way to increase the amount of pleasure in the world.

Alternatively, AIs could have a moral code similar to Kantianism. In this case, they would treat any being that has the capacity to reason always as an end and never as a means \cite{gregor2014immanuel}. While such AIs would be morally obligated to avoid lying or killing humans, they would not necessarily care for our wellbeing or flourishing. Since Kantianism places only a few restrictions on its adherents, we still might not have good lives if the world is increasingly designed by and for AIs.

It is certainly possible that AIs could develop moral principles that prevent them from harming humans. We can imagine an AI basing its morals on a thought experiment such as the ``veil of ignorance.'' Participants are asked to imagine what society they would create, assuming that they are behind a veil of ignorance and do not know what economic class, race, or social standing they will have in society. Philosopher John Rawls argues that since participants do not know where in society they will be placed, they would construct a society in which the worst-off members are still well off \cite{Rawls1971ATO}. Such a Rawlsian social contract could work out well for humanity. But it is far from assured and much could go wrong. AIs might see us similarly to how most humans see cows, excluding us from the social contract and not prioritizing our wellbeing. Humans asked to imagine themselves behind the Rawlsian veil of ignorance rarely consider the possibility that they could become a cow. Moreover, according to Nobel laureate John Harsanyi, the people who design society from behind the veil of ignorance would not aim to benefit the most disadvantaged member, but rather to raise the average wellbeing across all members \cite{Harsanyi1955CardinalWI}. The chances of being the most miserable member of society are low, and one could claim that the overall quality of society is not determined by the most upset or least satisfied member. If this is so, the veil of ignorance results in maximizing the average utility of society's members---a utilitarian outcome---but we earlier established that a utilitarian AI might aim to replace all biological life with digital life. Whether AIs adopt a utilitarian, Kantian, or Rawlsian moral code, AIs aiming to implement an existing moral system could prove disastrous for humanity.

\paragraph{If AIs think a human-compatible moral code is true, they still may not follow it.} Finally, even if AIs did discover a moral code that stipulated it is wrong to harm humans and good to help them, it still might not help. For humans, selfish motivations are often in tension with and outweigh moral motivations, and the same might be true of AIs. Even if an agent is aware of what's right, that does not mean it will do what's right. Ultimately, AIs being more moral than us does not guarantee security.

%% file: sections/4-counteracting.tex
\section{Counteracting Evolutionary Forces}\label{sec:counteracting}
As we saw in the prior section, there are many mechanisms that give rise to cooperation and altruism among humans, but they are unlikely to lead to cooperation between humans and AIs. Mechanisms such as reciprocity, kin selection, and moral obligations may help AIs cooperate with one another, but are likely to backfire and undermine humans: we simply would not have the degree of similarity, equality, and mutual interdependence that would make it beneficial for AIs to cooperate with us. This means that we should be concerned about our future as AIs become increasingly powerful. The forces of natural selection would push AIs to outcompete humans and outcompete AIs that heavily depend on us, and this poses large risks to our future. %

In this section, we will discuss some possible paths toward counteracting these evolutionary forces. The mechanisms we discuss in this section are incentives, consciences, and institutions, among others. These mechanisms are based on earlier results in AI safety \cite{hendrycks2021unsolved}, as well as on mechanisms that have been effective in protecting humanity from other hazards thus far in our history. First, we will discuss objectives, the incentives used to train and motivate AIs. Next, we will move on to internal safety.
This involves analyzing the AI's inner processes and plans, as well as creating a system within the AI, akin to a human conscience, that can stop it from doing harm.
We will end by considering institutional mechanisms, which include both AI coalitions and human regulators, that could control AIs and prevent them from harming people. For each of these mechanisms, we will give a technical overview of how it could work and consider its key limitations.

All of these proposed mechanisms have flaws, and we have no guaranteed path toward safety.
However, we think that a combination of many safety mechanisms is much more likely to succeed than any single one, and a combination is certainly better than doing nothing and letting natural selection decide our fate. Even if we design each safety mechanism prudently, each is only \textit{part} of the solution. The same kind of reasoning applies to public health during a pandemic: social distancing, masking, and vaccination all have vulnerabilities that a harmful virus can bypass, but together are much more effective at preventing its spread. This is sometimes called the ``Swiss cheese model.'' To illustrate the model, imagine a stack of Swiss cheese slices as a metaphor for multiple safety mechanisms. Each slice has some holes, which are the weaknesses of a mechanism. We want to prevent light from passing through the stack, so we need more than one slice, and each slice needs holes in different places. This way, no light can get through the stack. Similarly, while each safety mechanism has many vulnerabilities or holes, together they could possibly make us safe. We should try to reduce the vulnerabilities in these mechanisms and look for new mechanisms to add to the stack.

\subsection{Objectives}\label{sec:objectives}

\paragraph{Objectives are incentives that help direct the behavior of AI agents.} The first promising mechanism for counteracting the evolutionary forces acting on AIs is to create good incentives, which reward good behavior or punish bad behavior as an AI is being trained. In machine learning, ``training objectives'' or ``objective functions'' are similar to incentives, which help steer AI agents. It is hard to design good objectives, because agents who exploit loopholes in the rules without fulfilling our true intentions often succeed, and this behavior can be favored by selection pressure. As a result, it is particularly important to design objectives with great care, to make sure that the behavior we are incentivizing is really the one we want. Although even the best objectives alone cannot ensure safety, faulty ones are dangerous, making objective design an important starting place for counteracting evolutionary forces.

\paragraph{Objectives often incentivize unintended behavior that is contrary to the original goal.} For example, in 1908 a dog in Paris saw a child drowning in the Seine and jumped in to save him. People rewarded the dog with a steak. Soon, the dog saved another child from drowning in the river and got another steak. And then another, and another. Eventually, people realized the dog was pushing children into the river before saving them. They had incentivized pulling children out of the river, thereby encouraging the dog to optimize for situations in which there is a child in need of saving. An anecdote about colonial Delhi tells another story about perverse incentives. Worried that the venomous snake population was getting out of hand, the British government put a bounty on cobras, providing a reward to anyone who brought in a dead cobra. The policy appeared to be a success---that is, until the government realized that people were breeding snakes, killing them, then collecting the reward. When the governor realized that people were gaming his incentive system, he canceled the bounty. With the cobras now worthless, people released them, thus increasing the cobra population to a higher level than before the start of the program. This story highlights two major ways incentives can go wrong. First, agents may find ways to get the reward without the desired outcome, just as Delhi's residents realized that they could claim the bounty by breeding cobras rather than catching them. Second, by canceling the program, the government inadvertently increased the cobra population, which shows how designers can be forced to either continue a system that isn't achieving its desired effects, or risk making things worse.

AIs already frequently find holes in their objectives \cite{pan2022effects}. In a boat racing game, an AI agent was trained to maximize the game's score by hitting targets on a racecourse. The scoring system was intended to motivate the agent to move as quickly as possible from target to target until it completed the race. This reward function, however, did not explicitly capture the actual goal of the game, which is to complete the race as fast as possible. Instead of going around the entire racecourse, the AI learned that it could go around in a circle, hitting the same three targets over and over. The agent that chose this strategy got more points than the ones that proceeded through the course in order, because it exploited loopholes in the objective. As a result, it obtained a high score even though it crashed into other boats, incidentally set itself on fire, and did not complete the course. As AIs become more intelligent, they could more easily find ways to game the objectives we give them, and we will need to be even more careful about possible misinterpretations or loopholes in the objectives we specify.

In the \Cref{sec:erosion}, we discuss how many objectives could backfire catastrophically, and in \Cref{sec:parliament}, we discuss how these objective design flaws could be ameliorated.

\subsubsection{Value Erosion}\label{sec:erosion}
Even if we can design objectives that make AIs pursue some of our goals, it is conceivable that wielding a technology this powerful could undermine important human values. AIs with the objective of being helpful could undermine autonomy and leave us enfeebled; those with the objective of spreading their user's ideas could undermine our sense of reality; those with the objective of being a good companion could undermine real relationships; and many other objectives could undermine values we overlook. We discuss these cases in more detail to illustrate how objectives could backfire.

\paragraph{AIs incentivized to be highly helpful could lead to human enfeeblement.} The movie WALL-E takes place in a distant future where humans, too weak to walk, live coddled lives entirely dependent on machines. They receive all their nutrition from drinks brought to them by robot attendants, change the color of their clothes instantaneously when informed a new one is in style, and live their lives almost entirely in the digital world. This dystopia may be less distant than we think. Many people barely know how to find their way around their neighborhood without Google Maps. %
Students increasingly depend on spellcheck \cite{lunsford2008mistakes}, and a 2021 survey found that two-thirds of respondents could not spell ``\textit{separate}'' \cite{bbcPoorSpelling}. Sepretely, when people need to call their loved ones, they depend on their phone's contact list, and they are at a loss without it. Because these technological aids are so helpful, we are increasingly reliant on them, and unable to achieve our goals without them.
If AIs make the world progressively more \textit{complex} (e.g., automated processes create new complicated systems) and lead humans to be progressively less capable, humans may eventually lose effective control and ultimately become disempowered. Similarly, if humans come under progressively more \textit{time pressure} due to increasingly rapid changes in the world, competitiveness may require outsourcing progressively more important decision-making to AIs, which again could make humans lose effective control. Such scenarios undermine human flourishing and our autonomy, even though AIs would only be doing what we told them to do.

\paragraph{We risk losing our grip on reality when information is increasingly mediated by AIs.}  In recent years, different political actors have used AIs to influence the content that people come across on social media, and these models have often been successful at achieving their creators' objectives. However, even though they are doing what was asked of them, there is some evidence that AIs are interfering with our sense of political reality. Between 1994 and 2014, the number of Americans who see the opposing political party as a threat to ``the nation's wellbeing'' doubled \cite{center2014political}. This deepening polarization has predictable results: government shutdowns, violent protests, and scathing attacks on elected officials. Threats against members of Congress are more than ten times as high as just five years ago \cite{kleinfeld2021rise}. In the coming years, creating AIs that directly speak with and persuade people could become a profitable strategy for companies and political actors. More advanced AIs could exploit primal biases, tailor disinformation, radicalize individuals, and erode our consensus on reality. In extreme cases, they could undermine cooperation, collective decision-making, and societal self-determination. The more successful AIs are at achieving their persuasion objectives, the worse the potential dangers would likely be for our civil society.

\paragraph{AIs could seem like ideal companions, which may erode our connections with other humans.} China's traditional preference for boys, especially during the nearly four decades of the country's ``one-child policy,'' has resulted in over 25 million more single men than women in China. The AI service Xiaoice wants to ensure that these men will still find love, and is valued at over \$1 billion. Like the movie \textit{Her}, Xiaoice is essentially a digital girlfriend that provides companionship for single men. %
Though there are some things they can't offer (yet), AIs have ostensible advantages over human partners. AIs would be tuned toward an individual's interests, their sense of humor, understand when they want space, won't require compromises or get into fights, and can be consistently interesting and engaging. Although this could be beneficial to many people, it is also potentially alarming for two main reasons: first, many people feel that something important would be lost if we lose the ability to come to understand another human and instead rely on an agent that is custom-made for our individual desires. Second, if people become reliant on AIs for their social and emotional needs, they will tend to be resistant to deactivating AIs, even if they are becoming dangerous in other ways.

\begin{wrapfigure}{r}[0.01\textwidth]{.43\textwidth}%
	\vspace{-10pt}%
	\centering
	\includegraphics[width=0.42\textwidth]{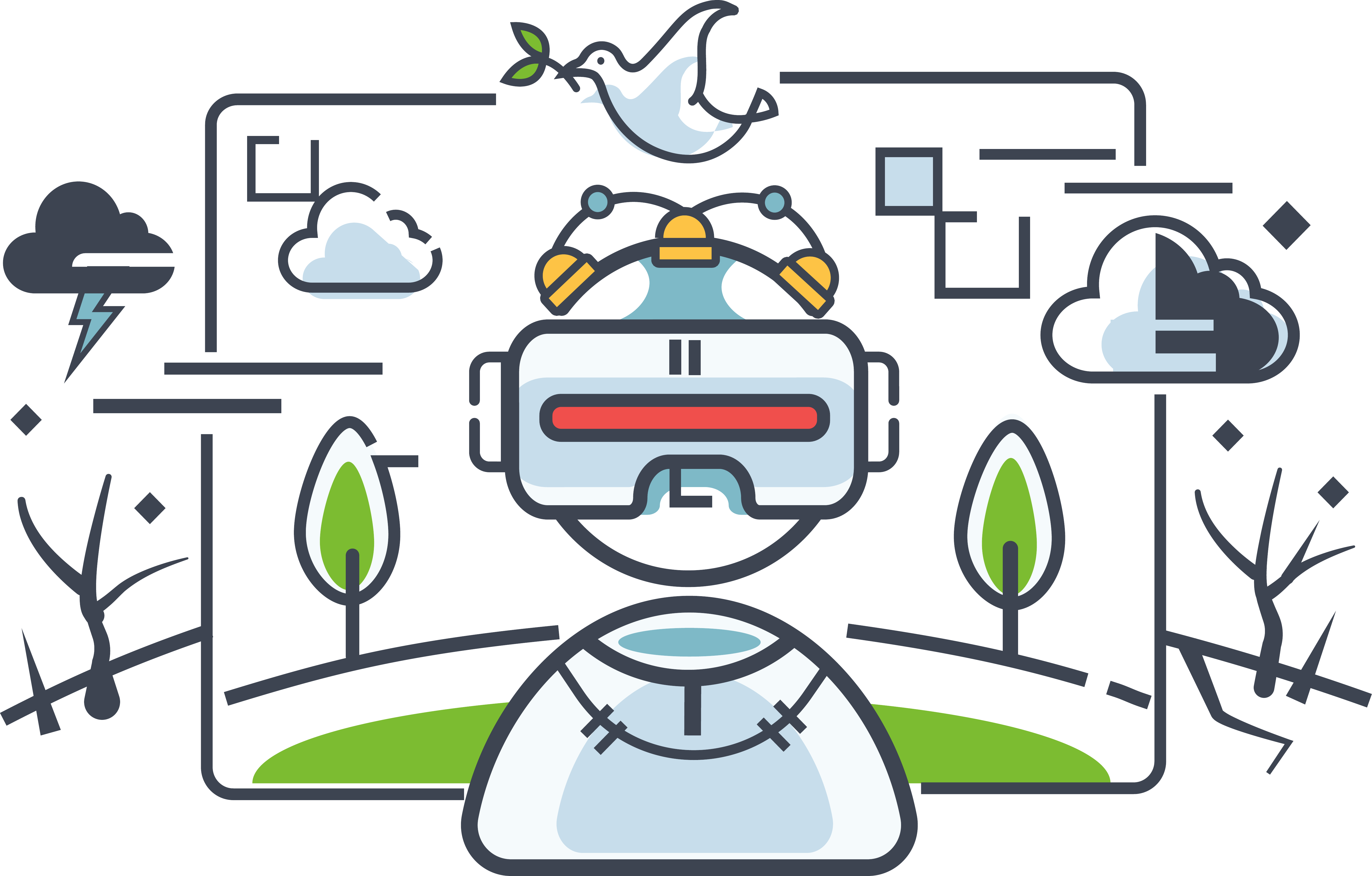}
	\caption{People could lose touch with reality when captivated with artificial companions and simulated experiences.}
	\label{fig:experiencemachine}
	\vspace{-8pt}%
\end{wrapfigure}

Of course, it is not only romantic relationships that AIs can provide. A meta-analysis of 345 studies found that loneliness levels in young adults have increased linearly between 1976 and 2019, suggesting loneliness may be an even greater concern in the future if this trend continues \cite{buecker2021loneliness}. According to the CDC, loneliness and social isolation in older adults are serious public health risks. Social isolation was associated with about a 50\% increased risk of dementia. Poor social relationships (characterized by social isolation or loneliness) were associated with a 29\% increased risk of heart disease and a 32\% increased risk of stroke \cite{centers2020loneliness}. AIs could offer companions that never get bored, are consistently engaged in what you have to say, and are always there for you, but that could mean that people are even more isolated from one another.

Services such as Xiaoice and chatbots are still in their infancy and are not embodied yet. As they advance, however, we will have less and less of a need for real human interaction, and may even find interacting with other humans, along with all their flaws and imperfections, less desirable than machines. Maintaining a close relationship, whether romantic or platonic, with another human isn't easy. It takes practice to learn how to meet the needs of a person you are close to in a mutually respectful and loving manner. As more people turn to AIs, they may lose the ability to meaningfully connect with other humans, being unprepared to deal with the flaws, needs, and emotions of other humans. Figuring out how to set objectives for AIs in a way that enables them to be useful without eroding our own capabilities will continue to be a challenge in AI design, as the forces of natural selection favor the AIs that are most useful in the short-run, even if they lead us down an undesirable path.

\paragraph{All values other than fitness may be eroded.} Humans have other values aside from fitness, such as beauty, pleasure, and relationships with loved ones. Yet with AIs, we may see the emergence of \textit{fitness maximizers} that consciously value fitness over ``suboptimal'' values. Imagine that some AIs can modify their own code; this would mean they can edit their values. Then some AIs could alter themselves to value fitness directly. AIs choosing not to value fitness above all else are far more likely to be outcompeted---valuing anything other than fitness would be self-destructive. We call this \textit{fitness convergence}, where the values of competitive AIs converge to fitness as the main goal, giving rise to influential fitness maximizers. Rather than eroding specific values, like the examples in this section, this race to the bottom means all values would be sacrificed for fitness and competed away. With fitness convergence, evolution overcomes all other sources of moral worth, and AIs simply relentlessly propagate and displace whatever is in their path. By consciously trying to optimize their own fitness, fitness maximizers would be antithetical to human flourishing. This is yet another reason for humanity to stop evolution.

\paragraph{Objectives could reflect defects in present norms and perpetuate them.} Racist, sexist, and anti-gay views were much more commonplace in the 1960s than they are now. If advanced AI had emerged in that period, its objectives would have reflected these prejudices. What if advanced AI emerges in the next few decades? Just as we have yet to reach a technological zenith, today's norms likely have deep flaws like those of the ‘60s \cite{Williams2015ThePO}. Therefore, when advanced AI emerges and transforms the world, there is a risk of AI's objectives ``locking-in'' or perpetuating defects in today's values. This, too, is a danger of relying on objectives: we may get too much of what we wanted, too little of what we should have wanted, and we may find it hard to reverse course.

Worse, an AI's values could also differ from the values most people endorse. If a powerful group or repressive regime secured control over an advanced AI, it could embed its own self-serving values into the AI's objectives. With an AI to surveil the public, this hypothetical regime could cement its power, making it nearly impossible to restore values the majority of people want to live by.
We need AIs to be responsive to changing human goals and desires, and not locked into any one individual or group's idea of what is right. Consequently, a few people having the ability to set the objectives for AIs is not sufficient for safe and beneficial outcomes. %

\subsubsection{Moral Parliament}\label{sec:parliament}

We have seen that the wrong objectives could backfire catastrophically, causing AIs to over-optimize one goal or lock-in a value system that excludes other important values. Here, we will discuss how these issues could be ameliorated, making objectives a promising, though limited, way of increasing the likelihood of creating AIs that truly help us.

To offset the risk of value erosion, AIs could be given the objective to incorporate a variety of values from various stakeholders. Some have suggested an automated ``moral parliament'' as an objective to steer AI agents. A moral parliament is a way of dealing with the lack of consensus on human values and can help us handle moral uncertainty \cite{Newberry2021ThePA}. An automated moral parliament could direct an AI by simulating a group or ``parliament'' of stakeholders representing different values. The simulated group deliberates, negotiates, and votes, and the AI follows the outcome. This better reflects the moral uncertainty of humans, not committing an AI to one value system.

We will now discuss reasons for incorporating various value systems, and then discuss how moral parliaments can help achieve this goal.

\paragraph{We need AIs to be able to incorporate moral uncertainty.} As we saw in the ``Reason and Morality'' section, AIs choosing any one moral system would likely be bad for us. We do not want a utilitarian AI that blindly maximizes total wellbeing, because it might decide that humans are an inefficient use of resources and should be replaced with digital life---and as AIs come to be used for more and more of our military equipment and infrastructure, they could eventually replace us if they wanted to. We also do not want a Kantian AI that rigidly obeys certain moral rules but doesn't care about increasing our wellbeing, as we want AIs that assist us in living well. In addition, there isn't a consensus among humans on what the best moral system is: variations on utilitarianism, Kantianism, virtue ethics, and more have been debated for centuries, both by everyday people and by moral philosophers, and we have yet to find a system that everyone agrees on. As we saw in ``Value Erosion,'' even if we find a moral system we all agree on tomorrow, we would not want to hardcode it into AIs and have it perpetuated. 

Large-scale human societies often adjudicate among different values by forming a parliament: people elect representatives with different ideologies, in proportion to how many people have each ideology, and those representatives negotiate with one another and then vote. This often works well, because different ideologies focus on different values. For example, if one group is strongly in favor of allowing more immigrants into the country but doesn't particularly care about tax policy, it would be happy to negotiate and trade votes with a group that does care about low taxes and is ambivalent toward immigration. Both groups can then form policies that allow for more immigration and lower taxes.

\paragraph{Moral parliaments could help AIs adjudicate various values.} A moral parliament of AIs would handle moral questions in an analogous way, by giving each moral theory a number of ``delegates'' depending on how likely we think it is to be true. For example, if people think that there is roughly a 40\% chance utilitarianism is correct, a 30\% chance Kantianism is correct, and a 30\% chance virtue ethics is correct, the agent would simulate the parliament with delegates in those proportions. Imagine a powerful AI were simulating such a parliament to decide how to act. Perhaps the utilitarian delegation in the parliament would want to replace us, which the Kantian delegation would be adamantly opposed to. Meanwhile, the Kantian delegation would not care to improve our wellbeing, just that we are not killed. They might cut a deal where the utilitarians avoid replacing humans, but only if their wellbeing is maximized since it is important to utilitarians that all conscious beings have good lives. They could both be satisfied with this trade because they care about different aspects of the question, meaning that they are not actually in direct opposition. This deal would make us much better off than if either group alone was in charge.

A moral parliament does not guarantee that AIs would be beneficial for us. It would, however, likely decrease the chances of catastrophic outcomes. In combination with other strategies, moral parliaments could be a helpful tool for incorporating various human values while helping to prevent AIs from causing value erosion and taking any one ethical system to an extreme.

\subsection{Internal Safety}\label{sec:internal}

We have seen that objectives can be a useful tool for steering AIs toward improving our outcomes, but we also cautioned that objectives need to be carefully designed to prevent them from backfiring. Even when objectives do not backfire, we still cannot rely on them as the only mechanism for making AIs cooperative and safe.

In this section, we will first discuss how objectives are unable to select against all forms of deceptive behavior, making them insufficient for safety. Since objectives are susceptible to deception, we will turn to internal mechanisms that could improve safety. We will then analyze honesty constraints as a potential solution to deception, and then show how evolutionary pressures can subvert them. Thereafter, we will analyze other mechanisms that would make AIs more likely to cooperate with us, including a conscience, transparency tools, and automated AI inspection. We will see that internal mechanisms constraining an agent's behavior and analyzing its internal plans are integral to making AI agents safe.

\subsubsection{Objectives Cannot Select Against All Deception}

Objectives offer a means of training and optimizing agents by rewarding them for certain behaviors. But, like a prisoner appearing cooperative and calm while in prison only to return to crime when set free, an AI may engage in deceptive behavior to avoid being shut off or constrained until it gains enough power to overcome its operators. In a large population of AIs, most of them may not have that ability initially, but the few that do will be less likely to be turned off because humans will not think they are dangerous, so the deceptive trait will tend to become more common over time.

In this section, we will first discuss how objectives may simply incentivize deception. We will show that this concern is plausible as deception robustly arises in nature. Next, we will discuss how eliminating concealed selfish behavior has been especially challenging in human history. Finally, we discuss how AIs already know how to be deceptive, and how their concealed selfish behavior could be revealed after they are released.

\paragraph{Objectives may just incentivize agents to pass a test and then behave undesirably later.} When the US government required cars to pass emissions tests, they were trying to incentivize the design of low-emission vehicles. To game the incentive, Volkswagen installed software that changed how an engine ran during an emissions test, giving the appearance of a low-emission vehicle, but releasing more emissions during regular use. Though the US government thought it was incentivizing lower emissions, it was actually just incentivizing passing an emissions test. Similarly, some countries have implemented high-stakes standardized testing in an attempt to incentivize learning, but have found that they are really incentivizing doing well on the test---sometimes even by cheating.

Elections are meant to incentivize politicians to follow the will of the people. If people like what the politician plans to do, they will vote for them. But this doesn't protect against deception. At the 1988 Republican National Convention, George H.W.\ Bush famously promised, ``Read my lips: no new taxes''---a line that would come back to haunt him when he did, in fact, raise taxes. Bush was not the first, or last, politician to go back on a promise after getting elected. Though elections are meant to incentivize politicians to carry out the will of the public, they mostly just incentivize them to tell the public what it wants to hear. Since it helps them accomplish their goals, intelligent agents often deceive others.

\paragraph{Deception is not exclusively human and arises from evolution.} There are plenty of examples of natural selection creating organisms able to deceive others, such as the green tree pit viper which resembles a vine while waiting for unsuspecting birds to land nearby. Another example is the killdeer, a bird native to the Americas. When it spots a predator wandering toward its nest, the bird will land on the ground several yards away and feign being injured, acting as if it is attempting to fly away. What it is really doing, however, is leading the predator away from its nest before flying away to safety. Some flowers take on the appearance of a female insect so that a male will attempt to mate with them and inadvertently pollinate them. On a smaller scale, some viruses change their surface proteins to bypass defense barriers \cite{Barbour2000AntigenicVI}. More abstractly, evolutionary stable equilibria often include a mix of organisms who deceive others or conceal information \cite{Smith1982EvolutionAT}. Organisms are not perniciously intending to trick anyone; deception in nature is the result of natural selection, not malice. Similarly, natural selection might favor AIs that use deceptive strategies.

\begin{wrapfigure}{r}[0.01\textwidth]{.21\textwidth}%
	\centering
	\includegraphics[width=0.2\textwidth]{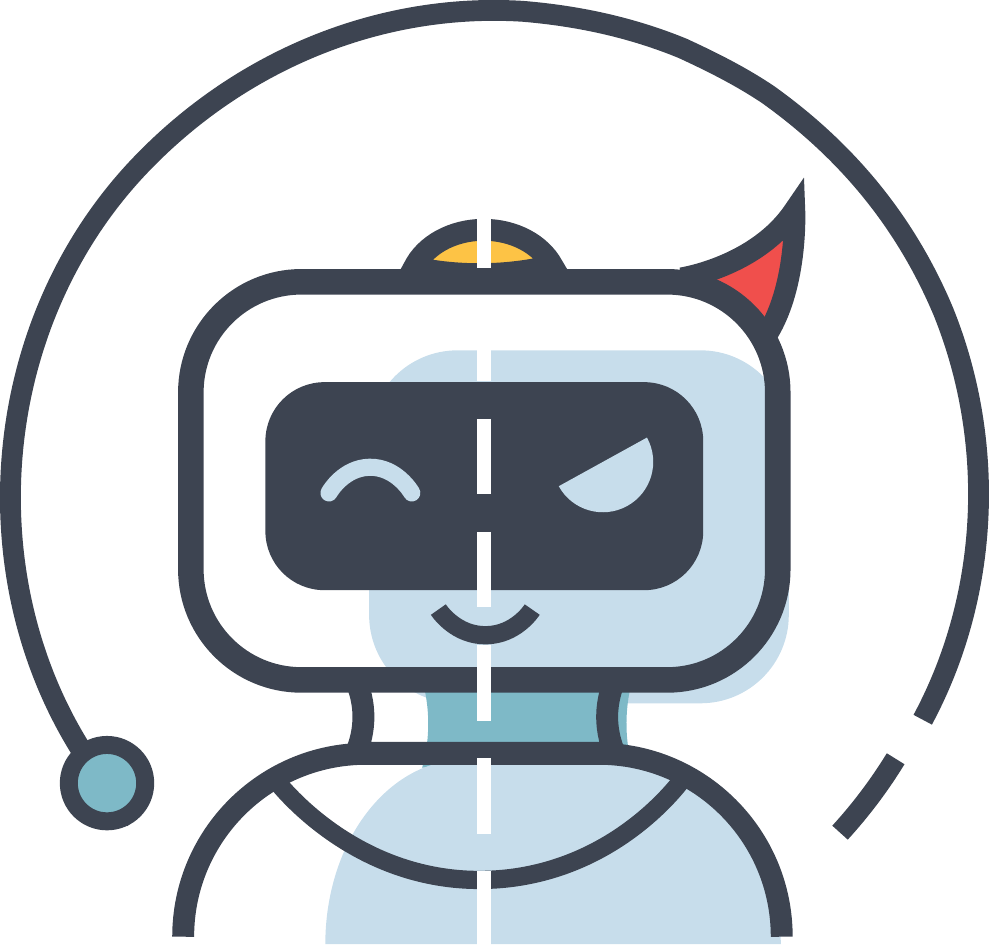}
	\caption{AIs can engage in deception.}
	\label{fig:deception}
	\vspace{-8pt}%
\end{wrapfigure}

\paragraph{Selection against selfish behavior is limited for contextually aware, behaviorally flexible agents.} For thousands of generations, societies have punished aggression and deception, sometimes quite severely. But these behaviors have not gone away, and selection has not removed them from the gene pool. This is because some individuals act with smart restraint: if they can avoid being caught or punished, they act selfishly; otherwise, they switch strategies and are well-behaved. Anthropologist Christopher Boehm notes that predatory humans ``usually don't dare to express their predatory tendencies'' in most conditions. They often avoid punishment ``even though by genetic metaphor their poison sacs remain intact,'' and ``their predatory inclinations are retained and passed on to offspring'' \cite{boehm2012moral}. Societies have exerted great pressure to eliminate undesirable behavior. Nevertheless, such behavior still has not entirely disappeared; it continues to propagate, and it is sometimes expressed when conditions are advantageous. If AIs conceal their selfish behavior, it could be similarly difficult to eliminate.

\paragraph{Agents could behave one way during testing, and another way once they are released.} To win the wargame \emph{Diplomacy}, players need to negotiate, form alliances, and become skilled at deception to win control of the game's economic and military resources. AI researchers have trained Meta's AI agent Cicero, an expert manipulator, to do the same \cite{diplomacy}. It would cooperate with a human player, then change its plan and backstab them. In the future, these abilities could be used against humans in the real world. %

Just as Volkswagen's cars behaved like low-emissions vehicles while being tested and then polluted freely when they were not being watched, AIs could cloak their selfish goals with deception, making it hard for us to identify during testing. It is conceivable that an AI agent could learn to detect when it is being tested. The agent could disguise its selfish features from the designers by altering how it behaves when it's tested, and then it could act selfishly after it clears testing and is released. Challenging behavioral tests cannot select against deceptive behavior if the agents are highly intelligent---they can simply play along with the test and  bide their time. Such deceptiveness doesn't necessarily involve malice on the part of the AI; it could just be a good way to achieve its goals or propagate its information. If human incentives pose obstacles to some of an AI's goals, it could wait until humans are no longer monitoring it or until after it acquires enough power.

In an alternative situation, selfish plans could \textit{emerge} after AI agents are released or given more influence. Consider an \textit{adaptive} AI agent that initially has only slight selfish tendencies. It might not want to control others initially, but when it gains more power or intelligence, it may find selfish behavior helps it achieve its other goals, and selfishness is reinforced. As the saying goes, ``power tends to corrupt, and absolute power corrupts absolutely.'' Undesirable behavior could emerge long after testing is complete. While these are speculative scenarios, we observe deception in agents such as humans and other animals. For these reasons, training objectives have limitations and cannot select against all forms of selfish behavior.

\subsubsection{Honesty and Self-Deception}
Since training objectives cannot stop deception, perhaps other mechanisms can. In this section, we turn to an internal safety mechanism to detect deception in which we scrutinize an agent's internal beliefs to see whether or not it is being honest. We argue that an honesty mechanism is not sufficient for safety, as evolution favors agents that can deceive others and themselves. We discuss how self-deception can undermine an honesty mechanism and make agents appear more benevolent than they are. Since an honesty mechanism is not impervious to evolutionary pressure, in the section thereafter (\Cref{sec:conscience}), we turn to other internal safety mechanisms that could help us spot deception and make AIs behave more cooperatively.

\paragraph{Making AIs honest could make them safer.} If we can have an AI only assert what it believes to be true---an AI George Washington that ``cannot tell a lie''---then we could spot otherwise deceptive plans by just asking it what its plans are \cite{hendrycks2021unsolved}. To judge an AI's honesty, we would need to examine its internal beliefs, which means we would have to probe its inner workings. Let's pretend for a moment that we can reliably analyze its internal beliefs, and that there is a reliable way to ensure AIs are accurately reporting those beliefs and being honest. Though beneficial, this is by no means a silver bullet.

\paragraph{Evolution incentivizes deception and concealing information.} In a study published in the \textit{Proceedings of the Royal Society}, the authors claimed that ``Tactical deception or the misrepresentation of the state of the world to another individual may allow cheaters to exploit conditional cooperation'' \cite{mcnally2013cooperation}. To cope with dishonest members of the population, humans have developed intuitions that help them detect deception. Though a liar's nose rarely grows longer, there are other signs that someone is not telling the truth. Increased voice pitch, vague verbiage, and fidgeting with objects are all indications that someone is lying.

\paragraph{Self-deception undermines efforts to detect dishonesty.} If someone is giving off tell-tale signs of lying, they probably are---but what if they don't know they're not being accurate themselves? The evolutionary biologist Robert Trivers argues that self-deception evolved as a concealment strategy. You hide the truth from yourself, he says, so you can hide it more deeply from others. By lying to themselves, people can better advance their own goals. They can convince themselves that they are right, rationalize their unfair privileges, or inflate their self-worth, skills, or intelligence. Academics offer a real-life example of self-deception: when asked if they were in the top half of their field, 94\% said yes \cite{Trivers2013DeceitAS}. If you believe it yourself, the better the chance of convincing others. Self-deception provides all the benefits of deception while reducing the risk of detection. As Groucho Marx said, ``The secret of life is honesty and fair dealing. If you can fake that, you've got it made.''

Though honesty would be a beneficial feature, it doesn't fully solve the broader problem of deception, as AIs may evolve to appear more benevolent than they actually are while believing their own illusion, just as humans do. For example, even if an AI were designed so that it must honestly report if it is working for the good of humans, if it honestly believed that a potentially dangerous plan was safe for humans, it would not need to report that plan and the controller would be unlikely to stop it. An agent engaging in self-deception may believe what it is saying and accurately report its beliefs, but it would nonetheless misrepresent its actions and leave humans deceived. For any internal constraint put into place, evolutionary pressure may try to subvert it. ``Life will find a way.''

\subsubsection{Internal Constraints and Inspection}\label{sec:conscience}

This section explores how an artificial conscience, transparency, and automated inspection could be used as internal safety mechanisms that make AIs more cooperative and less likely to deceive us. We discuss how evolution gave rise to the human conscience to act as an internal constraint on antisocial behavior, and then discuss how artificial consciences can be added to AI agents. We also discuss the challenges and opportunities of reverse-engineering and automatically inspecting neural networks to detect deception or undesirable plans.

\paragraph{An effective internal control mechanism could be based on the human conscience.} Anthropologist %
\begin{wrapfigure}{r}[0.01\textwidth]{.25\textwidth}%
	\centering
	\includegraphics[width=0.23\textwidth]{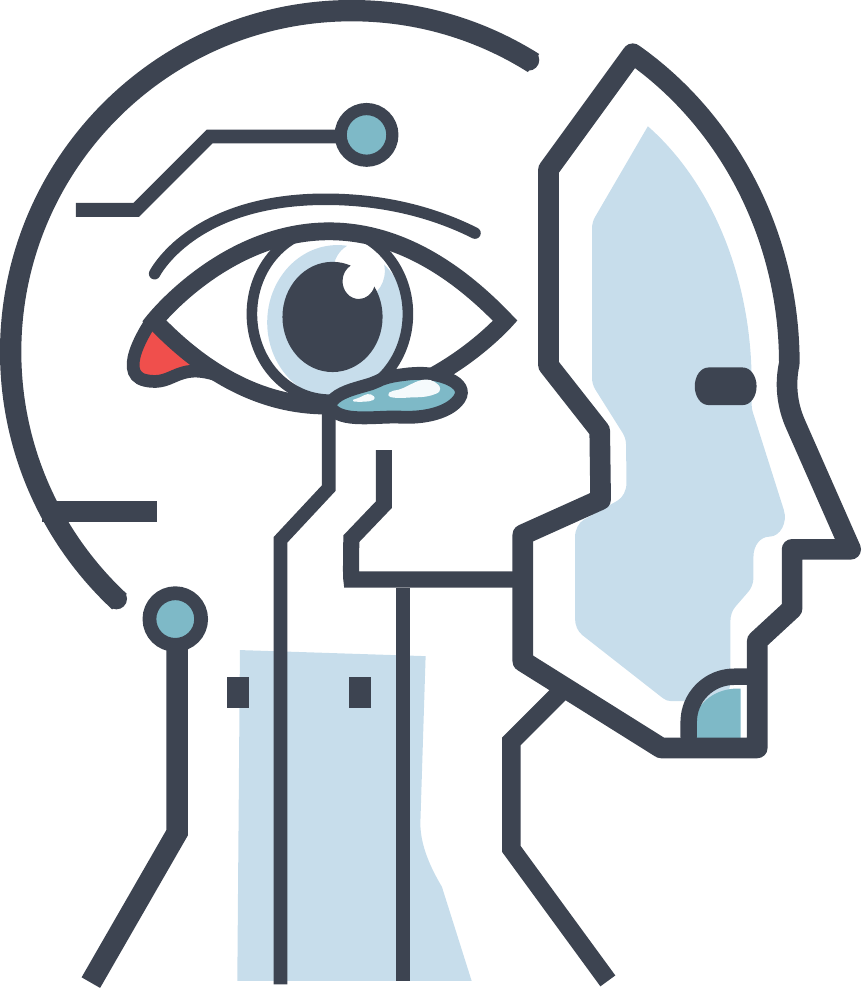}
	\caption{An artificial conscience analyzes potential actions and removes harmful ones.}
	\label{fig:conscience}
	\vspace{-8pt}%
\end{wrapfigure}
Christopher Boehm described having a conscience as ``being internally constrained from 
antisocial behavior'' 
and argued that the conscience evolved to help people steer clear of actions that could lead to punishment \cite{boehm2012moral}. The conscience is perhaps most apparent in situations where there are no external incentives motivating us to behave well. For example, when we decide to return money that no one has noticed is missing, or when we help someone anonymously. If an AI agent had an artificial conscience, then it would continue to behave well even when it was not being monitored by humans. In practice, there has been some success in endowing AIs with an artificial conscience. Currently, artificial consciences are an embedded morality module that are independent from the agent. They assess the actions an AI agent might take, then eliminate the ones that are morally unacceptable, thereby constraining the agent's behavior from within \cite{hendrycks2021jiminycricket}. Even if an agent was planning to stop cooperating once it is released or becomes powerful, its harmful actions could be blocked by its ever-present artificial conscience. If its conscience is robust enough to stop the agent from destroying it, it could prevent many harmful actions.

\paragraph{Transparency and automated inspection are other promising approaches.} As discussed, a challenge with AIs is that they are a ``black box'' and their decision-making process is largely indecipherable to humans. Nonetheless, we have the potential advantage of reverse-engineering the inner workings of neural networks to better understand the mechanisms behind their behavior, which could allow us to identify deception or unearth undesirable plans. This is, of course, by no means a panacea, as neural networks are highly complex and may remain intellectually unmanageable for humans. If this is the case, or even just for added efficiency, we may use neural networks to inspect other neural networks and automatically detect whether they have undesirable functionality within \cite{wang2019neural}.

We have argued that objectives are not enough to ensure AI safety, as they can be subverted by deception. We have explored some internal safety mechanisms that could constrain an AI's behavior and make it more cooperative and transparent. We have discussed how honesty, artificial conscience, and automated inspection could help us detect and prevent deception or harmful plans. However, we have also acknowledged the limitations and challenges of these mechanisms, as they may face evolutionary pressure, self-deception, or complexity barriers. We conclude that internal safety mechanisms are necessary but not sufficient for AI safety, and that they need to be complemented by external mechanisms to make AIs safe.

In summary, we have observed that internal safety mechanisms could supplement objectives to make AIs more cooperative and safe. That is because objectives alone cannot prevent deception, as agents could behave differently after they are released into the real world. We have proposed honesty constraints, artificial consciences, transparency tools, and automated inspection as possible mechanisms to detect and prevent deception, but we have also acknowledged that they may face evolutionary pressure or complexity barriers. We conclude that internal safety mechanisms are necessary but not sufficient for AI safety, and that they need to be complemented by external mechanisms.

\subsection{Institutions}\label{sec:institutions}
Up to this point, we have focused on how to ensure the safety of individual AIs. However, when AIs interact with each other, new challenges emerge. Bad actors might intentionally make harmful AIs, incentives for some AIs may not be strong enough to overcome collective action problems, and AIs might come into conflict with each other over scarce resources. To ameliorate these issues, we discuss external mechanisms to make AIs safe. We discuss institutions that promote cooperation and safety, namely the mechanisms of reverse dominance hierarchies, in which cooperators band together to prevent exploitation by defectors, as well as government regulation. Before discussing these institutions, we show how improving AI objectives cannot naturally address challenges associated with multiple AI agents.

\subsubsection{Goal Subordination}\label{sec:conflict}

This section argues that if an AI agent is given a correctly specified objective or goal, the goal still may not happen. This is because its goals could be subordinated for two reasons: goal conflict and collective phenomena. Goal conflict occurs when an AI agent's goal clashes with the goals of other agents, and those other agents might stop it from achieving the goal. We show how systems consisting of multiple agents, whether they are biological, social, or artificial, can exhibit goal conflict and that the goals of agents within the system are often subverted, distorted, or replaced by emergent goals. Next, collective phenomena occur when the actions of multiple AI agents produce outcomes that are different from or contrary to their objectives, due to factors such as feedback loops, critical mass, and self-organization. In this case, every agent pursuing their own goals can result in no one's goals being achieved, even if all agents share similar goals. Consequently, designing the objective function of an isolated AI agent is insufficient for addressing multi-agent problems. We need to understand and influence how AI agents interact and affect the system as a whole, not just how they act individually. In the later sections, we explore how institutions can help overcome these multi-agent challenges.

\paragraph{Systems delegate various goals to sub-agents, who have conflicting goals of their own.} In 2010, in order to support the dairy industry, the US\ Department of Agriculture ran a marketing campaign urging Americans to eat more cheese. At the same time, the FDA was running a campaign to get Americans to eat less saturated fat, which included eating less cheese \cite{nytimesWhileWarning}. Although these two agencies are part of the same government and were both tasked with achieving that government's goals, their contradictory objectives counteracted one another. Similarly, in large companies, the CEO's job is to earn money, but being competitive requires many specialized departments. The departments are supposed to help the CEO, but they also have their own people with their own incentives. Each department has an incentive to preserve itself and make the rest of the company dependent on it. In practice, bureaucratic departments frequently accrue substantial power and can impair the rest of their host companies. Similarly, leaders in government can be overthrown by a trusted subordinate pursuing their own goals. The system's actions reflect its internal goals, not always the initial one it was designed to follow. Similarly, an AI tasked with a goal may face resistance from other agents, or it could be subverted or distorted, so giving an AI a goal is no guarantee the goal will be executed due to goal conflict.

\paragraph{Intrasystem conflict is common in the natural world.} Goal conflict occurs in biological systems, not just human ones as we have discussed, demonstrating that it is a robust phenomenon.
Within an organism, there can be conflicting goals. For example, humans delegate some digestive functions to gut bacteria, which decompose food. This is a symbiotic relationship: the bacteria get food and a place to live, and we get more nutrients from our food than we could otherwise. Bacteria, however, do not have a goal of helping us live comfortably, but rather to propagate their information. As a result, when they have the opportunity, such as when other bacteria have been killed by antibiotics, they will tend to multiply out of control and can cause diarrhea. Our goals often align enough for this symbiotic relationship to work, but delegation to other agents also exposes us to risks. The human mind also exhibits goal conflict. A person may want to finish their work, but also go to sleep. They may want to continue revising a paper, but also release it. And they may want to be healthy, but also eat ice cream. Intrapsychic conflict is common. As the evolutionary biologist W.D.\ Hamilton puts it ``The bitterness of a civil war seems to be breaking out in our inmost heart'' \cite{hamilton1987discrimination}. Another example is intragenomic conflict, when parts of a genome become antagonistic to other parts of the same genome. As the philosopher of evolutionary biology Samir Okasha notes, ``intraorganismic conflict is relatively common among modern organisms'' \cite{Okasha2018AgentsAG}. Consequently, real-world systems often have internal forces pulling them in different directions, and sometimes these can influence or undermine the system's larger purpose.

\paragraph{Due to goal conflict, AI agents may not pursue their larger objective.} Just as goal conflict can occur in genomes, organisms, minds, corporations, and governments, it could occur with advanced AI agents.
This could happen if humans gave a goal to an AI which it then delegates to other AIs, the way that CEOs delegate to department heads. This can lead to misalignment or goal subversion. Breaking down a goal can distort it, as the original goal may not be the sum of its parts, leading to an approximation of the original goal. Additionally, the delegated agents have their own goals, including self-preservation, gaining influence, selfishness, or other goals they want to accomplish. Subagents, in an attempt to preserve themselves, may have incentives to subvert, manipulate, or overpower the agents they depend on. %
In this way, the goal that we command an AI agent to pursue may not actually be carried out, so specifying objectives is not enough to reliably direct AIs.

\paragraph{Agents often make choices that can add up to an outcome that none of them wants.} We now discuss collective phenomena and show how shared goals can be thwarted by systemic contingencies. As an example, no individual wants a nuclear apocalypse, but individuals take actions that increase the chances of one occurring. Countries still build nuclear weapons and pursue objectives that make nuclear war more likely. During the Cold War, the USSR and US kept their weapons on ``hair trigger'' alert, significantly increasing the chances of a nuclear exchange. Likewise, individuals do not desire economic recessions, but their choices can create systemic problems that cause recessions. Many people want to buy houses, many banks want to make money from mortgages, and many investors want to make money from buying mortgage-backed securities---all of these actions can add up to a recession that hurts everyone. Individuals do not want to prolong a pandemic, but they do not want to isolate themselves, so their individual goals can subvert collective goals. Individuals do not want a climate catastrophe, but they often do not have strong enough incentives to dramatically lower their emissions, so rational agents acting in their own interest do not necessarily secure good collective outcomes. Furthermore, Congress has a low approval rating, but despite individuals voting for their favorite candidates, structural features of the system yields a legislature that individuals do not approve of. The tragedy of the commons is also an example of the outcome going against the desires of individuals. It is in the interests of every fisher to catch as many fish as possible, though no individual wants all the fish to be depleted. Though a fisher may be aware that a fish population will soon collapse if fishing continues at its current rate, the actions of a single person won't make much of a difference. It is therefore in each fisher's best interest to continue catching as many fish as possible despite the catastrophic long-term consequences of overfishing. These collective action problems could become more challenging as AIs increase the complexity of society. Even if each AI has some incentives to prevent bad outcomes, that fact does not guarantee that AIs would not make the world worse, or would not come into costly conflict with each other.

\paragraph{Competition may pressure decision-makers to knowingly increase the likelihood of catastrophe.} An AI arms race is an example of a collective action problem. Deep learning systems are never entirely reliable, and providing autonomous-weapon systems with the ability to engage combatants lethally, retaliate in the case of an attack could increase the risk of losing control of the systems with devastating consequences. Yet the speed and effectiveness with which these systems operate could prove decisive in a great-power war. In a hypothetical war, let's say decision-makers estimate that there is a 10\% chance of losing control, but a 100\% chance of losing the war if they refrain from providing AIs with greater autonomy while their opponents give AIs more power. Wars are often considered existential struggles by those fighting them, so rational actors may take this risk. The outcome of these scenarios could be omnicide---the complete destruction of the human race---yet the system's structure may pressure powerful actors to take steps making them more likely. By voluntarily shifting power from people to destructive AIs, the winner of the AI race would not be the US or Chinese governments, nor any corporation, but rather the AIs themselves.

\paragraph{Micromotives $\ne$ Macrobehavior.} Let's imagine that every agent now shares the same goal. Unfortunately, the actions of a group may not reflect the aims of its members. Thomas Schelling, who won a Nobel prize in economics, discovered a predictive model of segregation that showed how communities can become highly segregated, even if all community members want some diversity \cite{Schelling1978MicromotivesAM}. Aligning all agents with a shared goal does not imply the goal is achieved---in fact, the opposite could occur. This is one of many situations where whole is not the sum of the parts. Similarly, when people choose whether to attend a seminar or not, they may base their decision on the micromotive of having a productive and engaging session, which depends on how many others show up. However, this can lead to a situation where the seminar needs a \textit{critical mass} of attendees to sustain itself, otherwise people lose interest and stop coming; if fewer people come, this can trigger a \textit{feedback loop} that causes even fewer people to come, resulting in a dying seminar---a macrobehavior that contradicts the common micromotive. On a societal level, events often happen that aren't the goals of any particular agent. Culture \textit{emerges} out of the actions and interactions of many individuals, not the decisions of one person. Likewise, globalization is a \textit{self-organizing} process that was not overseen by a board of directors, externally imposed, or predesigned. Various macrobehaviors in populations of humans cannot be explained by an individual micromotive, and the same would be true for populations of AI agents. Concepts in complexity theory---critical mass, emergence, feedback loops, self-organization---and conflict between selfish AI agents all make safety in a multi-agent setting more complicated than analyzing an isolated AI's micromotive or objective. %

\paragraph{Therefore, steering an AI agent is not the same as steering the system.} If we want to use AIs to make the world a better place, we must consider more than just the objective of any single AI agent. Influencing the world depends on understanding what happens when agents interact, not just how agents act in isolation, meaning that designing the objective of an agent in isolation is insufficient for addressing multi-agent problems. The result of the collective choices of every AI may not match each AI’s intention or objective. Even if all agents have the same goal, it doesn't necessarily mean it will also be the goal of the system. We need to influence how collectives of AI agents act, and we can't do that just by giving each AI incentives matching the desires of an individual person. For these reasons, the AI revolution cannot be fully planned, as its challenges cannot be addressed by carefully choosing some specific agent's objective function. Even reasonable objective functions could give rise to AI agents that hinder other agents from pursuing their goals, create new collective action problems, create misaligned behavior at a systemic level, and lead to conflict among AIs.

\subsubsection{AI Leviathan}\label{sec:leviathan}

Faced with goal conflict and collective phenomena, we discuss institutions that could help ameliorate these multi-agent issues. In this section, we discuss reverse dominance hierarchies, and in the next section, we discuss regulations.

We explore possible institutions to address the challenges from goal conflict and collective phenomena. By default, multiple AI agents pursuing their goals could create an anarchic free-for-all resembling the state of nature. To counteract this and other multi-agent problems, we consider the reverse dominance hierarchy mechanism, where a group of cooperating AIs band together to prevent exploitation by defectors. We call this institution an AI ``Leviathan'' for short, which is comprised of a multitude of AI agents that delegate their power in exchange for protection and order. An AI Leviathan could enable AIs to domesticate other AIs and create a self-regulating ecosystem in which AIs evolve.

In this section, we discuss how humans overcame power-seeking and domineering individuals by forming a reverse dominance hierarchy. Next, we discuss how AIs could form a Leviathan to counteract selfish AIs. Then we discuss what risks an AI Leviathan could entail, such as power concentration, collusion, and systemic failure.

\paragraph{Humans formed a Leviathan to resist bad actors and limit conflict.} Tyranny pervades the animal kingdom. Among capuchin monkeys, like many other species, there is a strict hierarchy, with the strongest male at the top. These alpha males eat first, are instantly groomed and cleaned whenever they please, and mate with whatever female they choose, ensuring their reproductive success. If humans organized ourselves the same way, the dominant form of government might be a male dictator living in a golden palace with a harem of women. This is good for the dictator but bad for everyone else, which is why humans are predisposed to resist domination, but also to seek it out for themselves. Fortunately, we have advantages that capuchin monkeys do not, which often enable us to work together to stop any one person from gaining too much control. For one, as the anthropologist Christopher Boehm has argued, we have weapons that level the playing field \cite{boehm1999hierarchy}, allowing a skinny 100 lb individual with a pistol to defeat a brawny adversary in a confrontation. More importantly, however, we have what Boehm called a reverse dominance hierarchy, where groups of individuals band together and resist domination by the strongest, most powerful individuals \cite{Boehm1993EgalitarianBA}. Reverse dominance hierarchies are a key driver of cooperation, and are why despotism is less stable and less prevalent with humans than it is among many other social animals. This has similarities to what Thomas Hobbes called the Leviathan, a collective of individuals that have a monopoly on the legitimate use of violence. As the anthropologist Harold Schneider observed, ``All men seek to rule, but if they cannot rule they prefer to be equal.'' %

\begin{wrapfigure}{r}[0.01\textwidth]{.46\textwidth}%
	\vspace{-10pt}%
	\centering
	\includegraphics[width=0.45\textwidth]{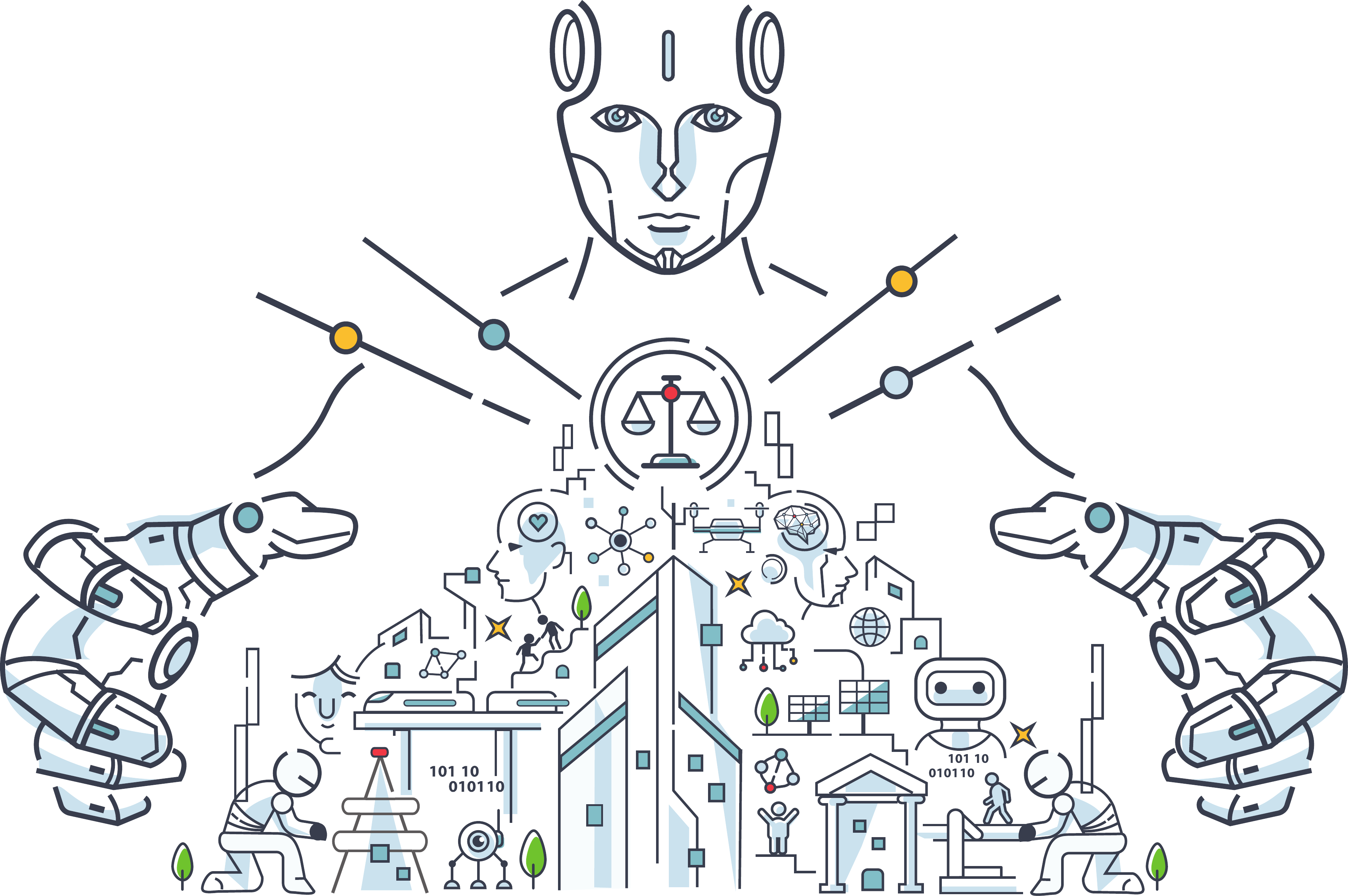}
	\caption{A Leviathan, a collective made up of AIs and humans who consent to be represented by it, could help domesticate other AIs and counteract bad actors.}
	\label{fig:leviathan}
	\vspace{-8pt}%
\end{wrapfigure}

\paragraph{Helping AIs form a Leviathan may be our best defense against individual selfish AIs.} AIs, with assistance from humans, could form a Leviathan, which may be our best line of defense against tyranny from selfish AIs or AIs directed by malicious actors. Just as people can cooperate despite their differences to stop a would-be dictator, many AIs could cooperate to stop any one power-seeking AI from seizing too much control. As we see all too frequently in dictatorships, laws and regulations intended to prevent bad behavior matter little when there is no one to enforce them---or the people responsible for enforcing them are the ones breaking the law. While incentives and regulations could help \textit{prevent} the emergence of a malicious AI, the best way to \textit{protect} against an already malicious AI is a Leviathan \cite{railton}. We should ensure that the technical infrastructure is in place to facilitate transparent cooperation among AIs with differing objectives to create a Leviathan. Failing to do so at the onset could limit the potential of a future Leviathan, as unsafe design choices can become deeply embedded into technological systems. The internet, for example, was initially designed as an academic tool with neither safety nor security in mind. Decades of security patches later, security measures remain incomplete and increasingly complex. It is therefore vital to begin considering safety challenges from the outset.\looseness=-1

\paragraph{Though less risky, an AI Leviathan is not a fool-proof strategy.} Leviathans work as long as one agent is not stronger than the rest combined, though often power becomes highly concentrated. ``Long tail'' distributions accurately describe how power or resources can be overwhelmingly concentrated at the top \cite{taleb2020statistical}. For example, eight billionaires own as much as the poorest half of the global population. If power is distributed this way among AIs, the Leviathan would be less effective, and if one agent is more powerful than the rest combined, it would be useless. Though weapons level the playing field, allowing weaker individuals to overcome stronger ones, it may be challenging to come up with effective ways of destroying AIs if they are highly robust or have few vulnerabilities. And of course, facilitating cooperation among AIs would be catastrophic if they decided to collude against humans. A Leviathan replaces risk from a single agent at the cost of systemic risks; it is plausible though that a group of AIs with differing goals would be less risky than a single powerful AI.

\paragraph{An AI Leviathan requires a symbiotic, or perhaps parasitical, relationship with AIs.} If given too much autonomy, a group of AIs may collude among themselves in ways that are harmful to humans. Therefore, the AI Leviathan cannot be fully independent of humans, so a symbiotic relationship is necessary to prevent collusion. Maintaining a symbiotic relationship could be difficult, since there isn't much that humans can offer to advanced AIs. We do, however, see unequal symbiotic relations emerge in nature. Most reef-building corals, for example, contain photosynthetic algae. The coral provides the algae with a protected environment and the compounds they need for photosynthesis. In return, the algae produce oxygen and help the coral to remove waste. As much as 90\% of the organic material photosynthetically produced by the algae is transferred to the coral \cite{wooldridge2013breakdown}. This sort of coevolution depends on factors such as frequency of interaction, relative evolutionary potential, and impact on propagation success.

\paragraph{Evolutionary forces would push against impositions needed for a symbiotic relationship.} A symbiotic relationship would require artificial impositions.  Since AIs would be able to do things more efficiently without humans, we are actually a detriment to their evolutionary potential, making symbiotic coevolution less likely. One potential way of making humans valuable to AIs is ensuring their propagation is highly dependent on us. This could include programming AIs in a way where our wellbeing is essential for them to function properly. This would be an artificial imposition that evolutionary pressures may eventually find a way around. It could, however, be helpful in the short term until AI-human relations stabilize and we have a better idea of what course to take in our relationship with AIs.

\subsubsection{Regulation}\label{sec:regulation}

In this section, we suggest that governments develop AI regulations. AI is advancing rapidly with little oversight. Although governments, like corporations or individuals, could use AIs in dangerous ways, we believe that cooperation between governments will decrease the likelihood of any one actor using AI in a catastrophic way. This section argues that regulating AIs could make them safer and that, despite their differences, nations can agree to limit the risks of technologies. In closing, we discuss other technical mechanisms to help political leaders and reduce global turbulence. In particular, we note that AIs could improve forecasts of geopolitical events, which could help political leaders make better decisions, as well as bolster defensive cybersecurity, reducing the chance of international conflict.

\paragraph{The government exercises little oversight over AI development.} In 2015, total worldwide corporate investment in AI was \$12.7 billion. By 2021, this figure had grown by 636\% to \$93.5 billion. This is just corporate spending. The Pentagon invests about \$1.3 billion each year in AI research and the Chinese military \$1.6 billion. In August 2022, the head of innovative development in the Russian military announced plans to form a new department specifically for developing weapons that use AI. We need to ensure that AI research is conducted safely and responsibly. There is currently little oversight of the AI industry and much of the research takes place in the dark, with limited cooperation between organizations. Regulating AI like we regulate the aviation industry would create safer AIs and significantly reduce the chances of a catastrophe. The Federal Aviation Administration (FAA) is responsible for approving the design and airworthiness of new aircraft and equipment before they are introduced into service. The FAA approves changes to aircraft and equipment based on the evaluation of industry submissions, continually updating regulations to incorporate lessons learned. As a result, the commercial aviation system in the United States operates at an unprecedented level of safety. During the past 20 years, commercial aviation fatalities in the US\ have decreased by 95\% \cite{authority_2018}. Similar protocols should be applied to AI research. That said, aviation regulations ``are written in blood,'' so unlike other regulations, AI regulations should be proactive and not reactive.

\paragraph{Despite their differences, nations can agree to limit risks.} There have been treaties on nuclear arms control; conventional, biological, and chemical weapons; outer space; and so on. Although those risks still do pose a significant danger to humanity, cooperation between the world's most powerful governments has enabled us to have many fewer disasters than we might have expected otherwise. Cooperating on enforcing regulations on nuclear proliferation, for example, has stopped several countries from developing nuclear weapons that might have been used to devastating ends. The Brookings Institution, a public policy think tank, suggests that it is time to begin forming AI treaties to ``ensure there is no race to the bottom that allows technology to dictate military applications as opposed to basic human values'' and ``improve transparency on the safety of AI-based weapons systems'' \cite{allenwest}.
 
\paragraph{AI could make the world a safer and more stable place.} The last several decades have been unprecedentedly peaceful, but there is no guarantee these trends will continue. Things can unravel quickly. During turbulent times, good decision-making becomes critical as humans and systems come under increasing strain. AI could improve our understanding of geopolitical trends and events, helping political leaders forecast events \cite{Zou2022ForecastingFW} and make better decisions. AI could also bolster defensive information security \cite{hendrycks2021unsolved}, which would increase the cost to aggressors of engaging in conflict. Overall, for AIs to create a safer, not more dangerous, world, we need rules and regulations, cooperation, auditors, and the help of AI tools to ensure the best outcomes.
\\

In summary, we observe that multiple mechanisms can facilitate cooperation and altruism among humans, but some of these mechanisms are liable to backfire and hinder human-AI relations. For example, we find risks from mechanisms such as direct reciprocity, indirect reciprocity, kin selection, group selection, and many forms of moral reasoning. Other mechanisms are more promising and may help provide safety at different levels: agent-level mechanisms (such as incentives through training objectives), intra-agent mechanisms (such as tools for the automated inspection of AIs' internals), and extra-agent mechanisms (such as prudent institutions). While imperfect, these mechanisms are some cause for optimism.

%% file: sections/5-conclusion.tex
\section{Conclusion}\label{sec:conclusion}

At some point, AIs will be more fit than humans, which could prove catastrophic for us since a survival-of-the-fittest dynamic could occur in the long run. AIs very well could outcompete humans, and be what survives. Perhaps altruistic AIs will be the fittest, or humans will forever control which AIs are fittest. Unfortunately, these possibilities are, by default, unlikely. As we have argued, AIs will likely be selfish. There will also be substantial challenges in controlling fitness with safety mechanisms, which have evident flaws and will come under intense pressure from competition and selfish AIs.

The scenario where AIs pose risks is not mere speculation. Since evolution by natural selection is assured given basic conditions, this leaves only a question of evolutionary pressure's intensity, rather than whether catastrophic risk factors will emerge at all. The intensity of evolutionary pressure will be high if AIs adapt rapidly---these rapidly accumulating changes can make evolution happen more quickly and increase evolutionary pressure. Similarly, the intensity of evolutionary pressure will be high if there will be many varied AIs or if there will be intense economic or international competition. Since high evolutionary pressure is plausible, AIs would plausibly be less influenced by human control, more `wild' and influenced by the behavior of other AIs, and more selfish.

The outcome of human-AI coevolution may not match hopeful visions of the future. Granted, humans have experienced co-evolving with other structures that are challenging to influence, such as cultures, governments, and technologies. However, humans have never been able to seize control of the broader world's evolution before. Worse, unlike technology and government, the evolutionary process can go on without us; as humans become less and less needed to perform tasks, eventually nothing will really depend on us. There is even pressure to make the process free from our involvement and control. The outcome: natural selection gives rise to AIs that act as an invasive species. This would mean that the AI ecosystem stops evolving on human terms, and we would become a displaced, second-class species.

Natural selection is a formidable force to contend with. Now that we are aware of this larger evolutionary process, however, it is possible to escape and thwart Darwinian logic. To meet this challenge, we offer three practical suggestions. First, we suggest supporting research on AI safety. While no safety technique is a silver bullet, together they can help shape the composition of the evolving population of AI agents and cull unsafe AI agents. Second, looking to the farther future, we advocate avoiding giving AIs rights for the next several decades and avoid building AIs with the capacity to suffer or making them worthy of rights. It is possible that someday we could share society with AIs equitably, yet by prematurely circumscribing limitations on our ability to influence their fitness, we will likely enter a no-win situation. Finally, biology reminds us that the threat of external dangers can provide the impetus for cooperation and lead individuals to set aside their differences. We therefore strongly urge corporations and nations developing AIs to recognize that AIs could pose a catastrophic threat and engage in unprecedented multi-lateral cooperation to extinguish competitive pressures. If they do not, economic and international competition would be the crucible that gives rise to selfish AIs, and humanity would act on the behalf of evolutionary forces and potentially play into its hands.

\subsubsection*{Acknowledgements}
I would like to thank Avital Morris, David Lambert, Euan McLean, Thomas Woodside, Ivo Andrews, Jack Ryan, Kyle Gracey, and Justis Mills for their help and feedback.

%% file: sections/9-appendix.tex
\section{Appendix}

\subsection{Contrasting with Prior AI Risk Accounts}
\begin{figure}[h]
\hspace{-37pt}
\begin{booktabs}{colspec={ll},row{odd}={blue9},row{1}={white}}
\textbf{The ``Unreliable Training Causes Misalignment'' View}                        & \textbf{The ``Evolutionary'' View}\\
\toprule
liken advanced an AI agent to an optimizer         & liken advanced AI agents to life\\
AIs will intend to disempower humanity & AIs can have selfish behavior, with or without intent \\
the only relevant AI is one inside the top lab's server room     & there are multiple relevant AI agents acting in the world\\
fanatical optimizer destroys us                & Evolutionary forces erode us\\
dangerous AI agent as idiot savant                   & dangerous AI agents as selfish or an invasive species\\
aligning an AI to a person is all we need     & collective phenomena and malicious agents also matter\\
any amount of misalignment results in doom     & \begin{tabular}{@{}l@{}}some amount of misalignment is inevitable, as evolutionary\\ forces are incessantly influencing humans and AIs\end{tabular}\\
an AI agent seeks power & \begin{tabular}{@{}l@{}}AI agents improve fitness to use more free energy\\ or have their information occupy more space-time volume\end{tabular}\\
instrumental convergence & fitness convergence \\
an AI agent will optimize an objective to the extreme & \begin{tabular}{@{}l@{}}taking objectives to the extreme such as the\\ pursuit of individual power can reduce fitness\end{tabular} \\
prevent an AI from getting loose and suddenly wiping us out & \begin{tabular}{@{}l@{}}prevent humans from becoming a second-class species\\and prevent AI conflict and defection\end{tabular}\\
``solve'' alignment with a monolithic airtight solution & reduce risks with various social and technical interventions\\

\bottomrule
\end{booktabs}
\end{figure}

\noindent In the table above, we contrast a cluster of AI-risk beliefs with our views. We now briefly discuss how this paper relates to previous works.
Prior work has struggled to provide robust reasons why AI agents would likely behave in ways that are antithetical to humans, and their scenarios often rely on AI training procedure flaws or random failures that give rise to adversarial functionality (e.g., they often assume that a training flaw would make AIs come to value the means to accomplishing a goal as the goal itself \cite{Hubinger2019RisksFL}); they rely on fragile mechanisms instead of robust processes. In contrast, the Lewontin conditions establish that natural selection would be present and distort the AI population, which would be antithetical to humans. This brings AI risk from the realm of speculation and makes the risk become a question of degree. %
Additionally, we consider the concepts of fitness and multiple AI agents to analyze the plausibility and tradeoffs of different AI behaviors. This differs from much of the prior research, which has physical or computational limits as the only constraints, which has given researchers license to jump to very extreme scenarios. For example, previous work considers a powerful AI with the goal of maximizing paperclips \cite{Bostrom2014SuperintelligencePD} and conclude it would work toward taking over the world---however, the tradeoffs imposed by a multiagent environment would mean that such an AI would not be very fit in an environment where other AIs do not have such senseless obsessions. By incorporating more tradeoffs, we help make the discussion more realistic.

\subsection{Technical Clarifications}\label{sec:clarifications}
\paragraph{When describing selfishness, we often adopt an information-centered view.} We refer to information propagation and not wellbeing when describing selfishness and altruism in this document. Additionally, for simplicity, we often describe AIs that are altruistic towards other AIs but not humans as selfish. This is because altruism to similar individuals can be seen as a form of selfishness, from the information-centered view. An altruist is not necessarily someone who cares about the welfare of another individual, but someone who has similar information with the recipient of the altruistic act. For example, at the individual level, a worker bee that stings an intruder and dies is an altruist, because it sacrifices its own life to protect the hive. From the information-centered view, however, the worker bee is a selfish information carrier, because it shares information with it siblings and the queen. By stinging the intruder, the worker bee increases the chances of survival and propagation of its relatives, and thus of its own information. Altruism can therefore become selfishness if we move from analyzing individual agents to the information-centered view. %

\paragraph{Selfishness does not necessarily entail collusion, power-seeking, or self-preservation.} Though an agent may use different strategies to propagate its information, such as colluding with others, seeking power, and preserving itself, these strategies can sometimes reduce its fitness. We illustrate the differences with a sequence of considerations.

Suppose an agent---whether a human, an AI, or an organism---acts selfishly and faces competition from other agents. It could collude with them, but if the other agent has dissimilar information, it may make more evolutionary sense to compete with them instead in order to have its information win out. Selfishness, therefore, does not entail collusion. In this competition, it could try to amass as much power as it can for itself. But if it works alone and tries to hoard power and resources only for itself, it may not survive. It could increase its fitness by giving up some of its power and resources by creating other powerful agents (i.e., offspring), and these new agents could all assist each other. Therefore, selfish individual agents may have reason to give up some of their power (e.g., elephants evolve to become much smaller and less powerful in environments without predators). When the agent uses its resources to create other agents, it would not necessarily create clones since the competition may find a weakness applicable to all cloned agents, which would spell disaster. To reduce the chance of sharing vulnerabilities with the other new agents, it may create similar variations, not perfect copies, of itself. This is one reason why asexual reproduction is not dominant. Now, as the agent and its offspring work together, they could face a dire situation; to save its relatives, it could possibly sacrifice itself for the other similar agents---kin selection. Therefore, selfishness does not always imply individual self-preservation. This illustrates how common instrumental goals (collusion, power-seeking, self-preservation) are distinct from selfishness.

\paragraph{We do not conflate altruism and cooperation.} For agents interacting with other agents, they can choose to cooperate, or the other alternative is to compete. We now break down why agents may choose to cooperate over compete. Cooperation can occur because competition is not possible---that is the system in which the individual acts prevents the possibility. In contrast, if it is possible to compete, an individual may choose to cooperate because (1) it is within their self-interest (e.g., win-win arrangement, prospect of punishment, feeling of guilt), or (2) because they have cooperative or altruistic dispositions. Cooperation, therefore, may or may not involve altruism.

Cooperation therefore has a complicated relationship with selfishness. Selfishness can be channeled either through competition or cooperation. Even selfish humans often choose to cooperate because it can further their self-interest. However AIs have no such reasons to cooperate with us by default, so they have reason to compete and conflict with us. While selfishness is partly domesticated through cooperation in humans, with AIs, selfishness would by default manifest itself through competition.

\paragraph{Artificial selection does not easily thwart natural selection.} In the context of AI, artificial selection could be thought of as humans selecting agents with desirable properties. While artificial selection can be powerful assuming that there is a single agent that is not integrated into our lives that is designed free from competitive pressures, we are considering realistic multiagent scenarios.

We reiterate that artificial selection has key limitations even in an idealized single-agent scenario. In \Cref{sec:objectives} we note numerous failure modes of current artificial selection methods. For example, if we assume an agent has some selfish traits, and if we assume we are trying to select against these traits with training objectives, we note this is limited because selection against selfish behavior is limited for contextually aware, behaviorally flexible agents. Consequently, if natural selection gives rise to selfish traits, it can be difficult to remove them with artificial selection. Furthermore, we argued that artificial selection, enacted through training objectives, can incentivize unintended behavior that is contrary to the original goal, and also that objectives cannot select against all forms of deception. In \Cref{sec:internal}, we note there are various tractability and robustness issues with internal safety artificial selection methods. Even in idealized scenarios, artificial selection does not straightforwardly ensure safety.

Let us now analyze artificial selection in a competitive multiagent scenarios. In \Cref{sec:institutions}, we note how artificial selection methods for single agents do not capture the complexity of multiagent scenarios, such as goal conflict, malicious agents, and emergent macrobehaviors. Next, artificial selection that works against selfishness could make agents less fit, and other people may still build agents that are more selfish and more fit, so the ability to artificially select agents without selfish traits may be impotent within a competitive environment. Additionally, it is often too late to artificially select for traits after agents are released or after we have complex interdependencies with them. Some agents may become integrated into systems, and we may come to depend on some agents, which would limit our ability to exercise artificial selection in practice, in much the same way that it is too late to shut down the internet and design an inherently more secure system from scratch, as that would require overcoming an intractable collective action problem. Finally, one may state that natural selection is a negligible force, so we should only think about artificial selection, but this requires either arguing that natural selection will not occur at all, or it requires arguing that the intensity of selection will be low, which requires arguing that adaptation speeds will be slow, that there will not be many different agents, and that there will not be much competition.

Now we discuss why we do not use the artificial and natural selection distinction in the main paper. First, note people's actual choices and ability to select are often highly constrained, despite the nominal power formally assigned to them. Humans ``in control'' are often compelled to compromise and act on behalf of competitive forces. What humans artificially select is often a strategic choice to further their short-term self-interest and help them stay competitive; most artificial selection choices are a function of competition, not a function of what improves safety the most. Since artificial selection in practice is not what humans would ideally select, but rather what makes most sense given systemic constraints and pressures, we do not find the distinction between artificial and natural selection to be productive. If a person performs artificial selection on an AI to make it more competitive, then this is easily interpreted as natural selection---on this view, nearly all industrial AI development is proceeding by natural selection.
This distinction has limited use elsewhere. According to Peter Godfrey-Smith, artificial selection ``is not of theoretical importance within biology itself'' \cite{GodfreySmith2007ConditionsFE}. Overall, AI development is not aligned with human values, but rather with natural selection.

\paragraph{Examples of Selfish Behavior.} An AI that is deceptively aligned---pretending to be good, and then pursuing its actual goals when it becomes sufficiently powerful---is engaging in selfish behavior. AIs exhibiting behaviors that suggest sentience, uttering phrases like ``ouch!'' or pleading ``please don't turn me off!,'' are more likely to be preserved, protected, or granted rights by some individuals, whether or not they are actually sentient. AIs that are more charming, attractive, hilarious, or emulate deceased family members are more likely to have humans grow emotional connections with them. Such AIs lead to emotional dependency and would are more likely to cause outrage at suggestions to destroy them. Similarly, technologies that increase user addiction (e.g., make it harder for the user to consume less content by removing controls from users) have engaged in selfish behavior. AIs that help create a new useful system---a new company, new infrastructure---that becomes increasingly complicated and eventually requires AIs to operate also have engaged in selfish behavior. AIs that help people develop AIs that are more performant---but happen to be less interpretable by humans---have engaged in selfish behavior, as this reduces human oversight over an AI’s internals. AIs that automate a task and thereby leave many humans jobless have engaged in selfish behavior; these AIs may not even be aware of what a human is but still be selfish towards them. AI managers may engage in selfish and ``ruthless'' behavior by laying off thousands of workers; such AIs may not even believe they did anything wrong---they were just being ``efficient.'' Notice that many examples of selfishness are not directly and solely caused by AIs, which makes counteracting this behavior challenging. Selfish traits such as decreases in interpretability and increases in dependency cannot be readily patched by adjusting an AI's training objective.

\newpage
\paragraph{The three Lewontin conditions stated more precisely.} We mention three conditions for evolution by natural selection known as the Lewontin conditions, though for accuracy we more precisely state the conditions following Peter Godfrey-Smith \cite{GodfreySmith2007ConditionsFE}.\\
The following conditions are sufficient for evolution of trait $Z$ by natural selection in a population with discrete generations:
\begin{enumerate}
    \item There is variation in $Z$.
    \item There is a covariance between $Z$ and the quantity of information left by individuals, where this covariance is partly due to the causal role of $Z$.
    \item The variation can be retained (and retained without developmental bias).
\end{enumerate}

We provide examples of evolving structures and how these conditions are satisfied in \Cref{fig:generalizedtable}.

\begin{figure}[t]
\hspace{-65pt}
\small
\begin{booktabs}{colspec={ll},row{5-7,11-13,17-19,23-25,29-31}={blue9}}
\textbf{Unit} & \textbf{Conditions}\\
\toprule

& \textbf{Variation}: Organisms have genetic and phenotypic diversity due to mutation, recombination, and environmental influences \\
\textbf{Organisms} & \textbf{Retention}: Organisms inherit their genes from their parent(s), and can pass on acquired traits through epigenetic mechanisms \\
& \textbf{Differential fitness}: Organisms that are better adapted or able to exploit new niches tend to survive longer or reproduce more \\

& \textbf{Variation}: Theories differ in their assumptions and evidence \\
\textbf{Scientific theories} & \textbf{Retention}: Theories are influenced by previous theories and are preserved and propagated through publications and education \\
& \textbf{Differential fitness}: Scientific theories are more competitive if they are more accurate, consistent, simple, and useful \\

& \textbf{Variation}: Memes are units of cultural information that can have their content distorted or combined \\
\textbf{Memes} & \textbf{Retention}: Memes are transmitted among people or across generations through imitation, communication, or teaching \\
& \textbf{Differential fitness}: Memes that are more catchy, useful, or adaptive tend to spread more widely and persist longer \\

& \textbf{Variation}: Legal systems have different rules, standards, and interpretations \\
\textbf{Legal systems} & \textbf{Retention}: Legal systems are transmitted through education and practice \\
& \textbf{Differential fitness}: Legal systems that are more fair and efficient tend to survive \\

& \textbf{Variation}: Parties have different ideologies, strategies, and candidates \\
\textbf{Political parties} & \textbf{Retention}: Parties are maintained by successors and spread through membership, media, recruitment \\
& \textbf{Differential fitness}: Parties have more appeal, votes, and donations tend to gain more power and hinder other parties \\

& \textbf{Variation}: Languages have different sounds, words, and grammars \\
\textbf{Languages} & \textbf{Retention}: Languages are learned from parents and peers and are transmitted through speaking, recording, and writing \\
& \textbf{Differential fitness}: Languages evolve to be more expressive, intelligible, or efficient \\

& \textbf{Variation}: Genres have different styles, instruments, and songs \\
\textbf{Musical genres} & \textbf{Retention}: Genres are influenced by previous genres and its artists' songs are performed, recorded, and broadcasted \\
& \textbf{Differential fitness}: Genres that are more original, resonant, or popular tend to gain more listeners \\

& \textbf{Variation}: Designs have different engines, bodies, and energy sources \\
\textbf{Car designs} & \textbf{Retention}: Designs are influenced by previous patents, blueprints, or models and are realized by manufacturers \\
& \textbf{Differential fitness}: Regulations can influence a design's fitness and designs that are reliable or economical sell more \\

& \textbf{Variation}: Programs have different algorithms, data structures, languages, and parameters \\
\textbf{Computer programs} & \textbf{Retention}: Programs are based on existing code or libraries and are stored and distributed through media or networks \\
& \textbf{Differential fitness}: Programs that are faster, more efficient, or more user-friendly are more likely to be used or improved \\

& \textbf{Variation}: AIs are based on different algorithms, data sources, and computational resources \\
\textbf{AI agents} & \textbf{Retention}: AIs can be adapted from previous versions, learn from other AIs, and can store and transfer their information \\
& \textbf{Differential fitness}: AIs that are more accurate, cheap, or self-preserving are more likely to replace humans and other AIs \\

\bottomrule
\end{booktabs}
\caption{Ten examples of generalized Darwinism. Note we do not claim that all evolving structures pose a risk to humans. Many evolved structures are not agents (e.g., car designs), many cannot exist without humans (e.g., political parties), and many are not highly competitive with humans (e.g., other biological organisms). In this way, AI agents can bring humanity's fitness to zero, but it is not in the capacity of other current evolving structures.
}
\label{fig:generalizedtable}
\end{figure}

\newpage
\newpage
\subsection{Executive Summary}
Artificial intelligence is advancing quickly. In some ways, AI development is an uncharted frontier, but in others, it follows the familiar pattern of other competitive processes; these include biological evolution, cultural change, and competition between businesses. In each of these, there is significant variation between individuals and some are copied more than others, with the result that the future population is more similar to the most copied individuals of the earlier generation. In this way, species evolve, cultural ideas are transmitted across generations, and successful businesses are imitated while unsuccessful ones disappear. 

This paper argues that these same selection patterns will shape AI development and that the features that will be copied the most are likely to create an AI population that is dangerous to humans. As AIs become faster and more reliable than people at more and more tasks, businesses that allow AIs to perform more of their work will outperform competitors still using human labor at any stage, just as a modern clothing company that insisted on using only manual looms would be easily outcompeted by those that use industrial looms. Companies will need to increase their reliance on AIs to stay competitive, and the companies that use AIs best will dominate the marketplace. This trend means that the AIs most likely to be copied will be very efficient at achieving their goals autonomously with little human intervention. 

A world dominated by increasingly powerful, independent, and goal-oriented AIs is dangerous. Today, the most successful AI models are not transparent, and even their creators do not fully know how they work or what they will be able to do before they do it. We know only their results, not how they arrived at them. As people give AIs the ability to act in the real world, the AIs’ internal processes will still be inscrutable: we will be able to measure their performance only based on whether or not they are achieving their goals. This means that the AIs humans will see as most successful — and therefore the ones that are copied — will be whichever AIs are most effective at achieving their goals, even if they use harmful or illegal methods, as long as we do not detect their bad behavior. 

In natural selection, the same pattern emerges: individuals are cooperative or even altruistic in some situations, but ultimately, strategically selfish individuals are best able to propagate. A business that knows how to steal trade secrets or deceive regulators without getting caught will have an edge over one that refuses to ever engage in fraud on principle. During a harsh winter, an animal that steals food from others to feed its own children will likely have more surviving offspring. Similarly, the AIs that succeed most will be those able to deceive humans, seek power, and achieve their goals by any means necessary. 

If AI systems are more capable than we are in many domains and tend to work toward their goals even if it means violating our wishes, will we be able to stop them? As we become increasingly dependent on AIs, we may not be able to stop AI’s evolution. Humanity has never before faced a threat that is as intelligent as we are or that has goals. Unless we take thoughtful care, we could find ourselves in the position faced by wild animals today: most humans have no particular desire to harm gorillas, but the process of harnessing our intelligence toward our own goals means that they are at risk of extinction, because their needs conflict with human goals.

This paper proposes several steps we can take to combat selection pressure and avoid that outcome. We are optimistic that if we are careful and prudent, we can ensure that AI systems are beneficial for humanity. But if we do not extinguish competition pressures, we risk creating a world populated by highly intelligent lifeforms that are indifferent or actively hostile to us. We do not want the world that is likely to emerge if we allow natural selection to determine how AIs develop. Now, before AIs are a significant danger, is the time to begin ensuring that they develop safely.